\documentclass[showpacs,superscriptaddress,amsfonts,amssymb,amsmath]{revtex4}
\usepackage{graphicx}
\usepackage{amsmath}
\usepackage{amssymb}
\usepackage{dsfont}

\newcommand{\csch}{\mathrm{csch}}
\usepackage{color}

\begin{document}

\title{Exact Solutions of a One-dimensional Quantum Spin Chain with $SO(5)$-Symmetry}
\author{Yuzhu Jiang}
\author{Junpeng Cao}
\author{Yupeng Wang}
\affiliation{Beijing National Laboratory for Condensed Matter
Physics, Institute of Physics, Chinese Academy of Sciences, Beijing
100190, People's Republic of China}

\begin{abstract}
A new exactly solvable one-dimensional spin-$3/2$ Heisenberg model
with $SO(5)$-invariance is proposed. The eigenvalues and Bethe
ansatz equations of the model are obtained by using the nested
algebraic Bethe ansatz approach. Several exotic elementary
excitations in the antiferromagnetic region such as neutral spinon
with zero spin, heavy spinon with spin-$3/2$ and dressed spinon with
spin-$1/2$ are found.
\end{abstract}

\pacs{75.10.Jm  
      02.30.Ik 
     }

\maketitle
{\small \vspace{-0.7cm}\hspace{1.18cm}Key words:
Exact solutions, Quantum spin chain, Algebra
Bethe ansatz}

\section{Introduction}

The one-dimensional(1D) quantum spin chains play a
very important role in the strongly correlated systems
and low-dimensional quantum magnetism and show many
interesting behaviors. For example, the ground state
of the spin-1/2 Heisenberg chain with
antiferromagnetic couplings is expected to be a spin
liquid rather than N\'eel ordered state due to the
strong quantum fluctuations. The elementary
excitations of such a system are usually described by
the spinons which carry spin-1/2 rather than spin
waves. In another hand, Haldane conjectured that the
spin systems with half-integer spins have  gapless
excitation spectra, while those with integer spins
have gapped spectra \cite{Haldane1983PRL}. The
spin-1/2 Heisenberg model was exactly solved by Bethe
\cite{Bethe1931}. By using Bethe's hypothesis, Yang
and Yang solved the $XXZ$ Heisenberg chain
successfully\cite{Yang1966}. Subsequently, Takhtajian
and Faddeev developed the algebraic Bethe ansatz
method and several spin chain models have been exactly
solved\cite{Takhatajan1982, Babujian1982,
Babudjian1983, Cao2007, Jiang2009}.

Recently, much attention has been focused on the high spin systems
because not only many peculiar quantum orders and exotic collective
excitations can appear in this kind of systems, but also some
materials such as CsVBr$_3$, CsVCl$_3$ \cite{Itoh1995}, CsVI$_3$ and
AgCrP$_2$S$_6$ \cite{Mutka1995} in nature can be modeled by the
spin-3/2 chain quite well. With the developments of experimental
technique of laser cooling and magnetic traps, atoms with high
nuclear spin such as $^{87}$Rb and $^{23}$Na with spin-1,
$^{132}$Cs, $^9$Be, $^{135}$Ba, $^{137}$Ba, and $^{53}$Cr with spin-3/2 can be
trapped. Using the Feshbach resonance techniques, one can tune the
scattering lengths among the atoms, which make it possible to
simulate the traditional solid systems with various interactions.
These progress provide us a ideal platform to study the physics in
high spin systems and many interesting quantum phenomena are found
\cite{Ziman1987, Alcaraz1992, HHTuGMZhangTXiang2008PRB78-094404}.

It is well-known that the spin-$s$ chain with the
$SU(2s+1)$-symmetry can be solved exactly, for the Hamiltonian can
be mapped onto the summation of permutation operators. Takhatajan
and Babudjian found that besides the $SU(3)$ integrable point, the
spin-1 system can still be solved at a special $SU(2)$-invariant
point\cite{Takhatajan1982, Babudjian1983}. They also showed that the
elementary excitation of the system is gapless. This motivate us to
seek other integrable points in the high spin systems. In this
paper, we show that besides the $SU(4)$ integrable point, there is
another exactly solvable model of the spin-3/2 chain with $SO(5)$
symmetry. By using the nested algebraic Bethe ansatz method, we
obtain the exact solutions of the model. Based on the exact
solutions, several exotic excitations such as the heavy spinon with
fractional spin 3/2, the neutral spinon with spin zero and the
dressed spinon with spin 1/2 are found, which are quite different
from those of the $SU(4)$ integrable spin chain.

The paper is organized as follows. We introduce the model and its
symmetry in Sec. \ref{M}. The nested algebraic Bethe ansatz approach
for the model is shown in Sec. \ref{ABa}. The thermodynamic
properties of the system are analyzed in Sec. \ref{TD}. The
ferromagnetic and antiferromagnetic ground states are discussed in
Sec. \ref{GS}. The elementary excitations are given in Sec.
\ref{Exc}. Sec. \ref{C} is a brief summary.

\section{The model}
\label{M}

As mentioned above, two integrable models of 1D spin-3/2 Heisenberg
chains, i.e., the $SU(2)$-invariant one and the $SU(4)$-invariant
one, have been found and solved. In this paper, we introduce another
integrable spin-3/2 chain model with $SO(5)$-symmetry. Our model
Hamiltonian reads
\begin{eqnarray}\label{HSS}
H = J\sum_{i=1}^N \left[\frac{25}{8} \vec S_i\cdot \vec S_{i+1}
-\frac73 \left(\vec S_i\cdot \vec S_{i+1}\right)^2 -\frac23
\left(\vec S_i \cdot \vec S_{i+1}\right)^3\right],
\end{eqnarray}
where $J$ is a coupling constant; $\vec S_i$ is the spin-3/2 operator at site
$i$, $i=1,2,\cdots, N$ and $N$ is the length of the system. Here, we
adopt the periodic boundary condition, i.e., $\vec S_{N+1}=\vec
S_{1}$.

To show the $SO(5)$ symmetry of our model, we introduce the
following Dirac matrices
\begin{eqnarray}
&&\Gamma^{1}=-\frac{1}{\sqrt3}(S_xS_y+S_yS_x), ~
\Gamma^{2}=\frac{1}{\sqrt3}(S_zS_x+S_xS_z), ~
\Gamma^{3}=\frac{1}{\sqrt3}(S_zS_y+S_yS_z),\nonumber\\
&&\Gamma^{4}=S_z^2-5/4,~\Gamma^{5}=\frac{1}{\sqrt3}(S_x^2-S_y^2).
\end{eqnarray}
The 10 generators of the $SO(5)$ Lie algebra can be expressed by
the Dirac matrices as $\Gamma^{a,b}=-\frac i2[\Gamma^{a},
\Gamma^{b}]$. The explicit from of these generators are \cite{CWuJPHuZShu2003-186402}
\begin{eqnarray}
&&\Gamma^{1,(2,3,4)} = \left(\begin{array}{cc}
0&\vec \sigma\\{} \vec \sigma &0\end{array}\right),~
\Gamma^{1,5} = \left(\begin{array}{cc}
-I&0\\{}0&I\end{array}\right),\nonumber \\
&& \Gamma^{3,4;4,2;2,3} = \left(\begin{array}{cc} \vec \sigma
&0\\{}0& \vec \sigma\end{array}\right),~ \Gamma^{2,(3,4,5)} =
\left(\begin{array}{cc} 0& -i \vec \sigma \\{}i \vec \sigma &
0\end{array}\right). \label{gg}
\end{eqnarray}
After some algebra, we find that the Hamiltonian (1)
commutes with the generators (\ref{gg}), $[H,
\Gamma^{a,b}]=0$. Thus the model has the $SO(5)$
symmetry. Different from the $SU(4)$ integrable spin
chain, there only exist three conserved quantities and
the number of spins with individual components is no
longer conserved. After some detailed analysis, we
find that the following quantities are conserved
\begin{eqnarray}
&&J_1=N_{3/2}+N_{1/2}+N_{-1/2}+N_{-3/2},\nonumber\\
&&J_2=N_{3/2}-N_{-3/2},\\
&&J_3=N_{1/2}-N_{-1/2}.\nonumber
\end{eqnarray}

The $R$-matrix of this model reads
\begin{eqnarray}\label{Rm}
R_{ab}(\lambda) = -\frac{2\lambda+3i}{2\lambda-3i}
P^0_{ab} + P^1_{ab} - \frac{2\lambda+i} {2\lambda-i}
P^2_{ab} + P^3_{ab},
\end{eqnarray}
where $\lambda$ is the spectral parameter; $P^{s}_{ab}$ is the
projection operator in the total spin-$s$ channel and acts on the
two coupled spin space $V_a\otimes V_b$.  Just as the $SU(4)$ one,
the non-trivial scattering processes only exist in the total
spin-$0$ and $2$ channels, but here the scattering strengths are
different in these two channels.

This $R$-matrix satisfies the Yang-Baxter
equation \cite{Jiang2009,Controzzi12006}
\begin{eqnarray}\label{YBERRR}
R_{ab}(\lambda)  R_{bc}(\lambda+\mu) R_{ab}(\mu)
&=& R_{bc}(\mu) R_{ab}(\lambda+\mu) R_{bc}(\lambda).
\end{eqnarray}
In the frame work of quantum inverse scattering method(QISM), the
Lax operators of the system are
\begin{eqnarray}\label{Lm}
L_{n}(\lambda) = R_{0n}P_{0n}=
\frac{2\lambda+3i}{2\lambda-3i} P^0_{ab}
+ P^1_{ab}+
\frac{2\lambda+i}{2\lambda-i} P^2_{ab} + P^3_{ab},
\end{eqnarray}
where $V_0$ is an auxiliary space and $V_n$ is the quantum space.
$P_{0n}$ is the permutation operator, which can be expressed by the
projection operators as
$P_{0n}=-P^{0}_{0n}+P^{1}_{0n}-P^{2}_{0n}+P^{3}_{0n}$. The monodromy
matrix $T$ is constructed by the Lax operators as
\begin{eqnarray}\label{monodromy}
T(\lambda) = L_N(\lambda)L_{N-1}(\lambda) \cdots L_1(\lambda).
\end{eqnarray}
The monodromy matrix $T$ satisfies the Yang-Baxter relation
\begin{eqnarray}\label{YBRRTT}
R(\lambda-\mu) T(\lambda)\otimes T(\mu)
=T(\mu)\otimes T(\lambda) R(\lambda-\mu),
\end{eqnarray}
Taking trace in the auxiliary space of $T$, we obtain the transfer matrix
\begin{eqnarray}
\label{transfer}
t(\lambda) = \mathrm{tr}_0 T(\lambda).
\end{eqnarray}
From the Yang-Baxter relation, one can prove that the transfer
matrices with different spectral parameters commutate with each
other,
\begin{eqnarray}
[t(\lambda), t(\mu)]=0.
\end{eqnarray}
These transfer matrices are the infinite conserved
quantities of this system. Thus the system is integrable.

Taking the derivative of the logarithm of the transfer
matrix, we arrive at the Hamiltonian (1)
\cite{Sklyanin1980, deVega1984NPB}
\begin{eqnarray}\label{HS-SP}
H = -\frac{9J}{4} \left. \frac{\partial \ln
T(\lambda)}{\partial\lambda} \right|_{\lambda=0}
-\frac{99}{8} JN= -3J
\left(P^0_{i,i+1}+3P^2_{i,i+1}\right) -\frac{99}{8}
JN,
\end{eqnarray}
Here, we have used the fact that the Lax operator (\ref{Lm})
degenerates into the permutation operator when the spectral
parameter is zero, $L_{ab}(\lambda)|_{\lambda=0}=P_{ab}$, and put a
constant $J$ into the Hamiltonian. The eigenvalue problem of the
Hamiltonian is therefore turned into the diagonalization of transfer
matrices. Suppose the eigenvalues of the Hamiltonian and the
transfer matrix are $E$ and $\Lambda$, then the eigenvalue $E$ can
be determined by the eigenvalue $\lambda$ as
\begin{eqnarray}
E = -\frac94J  \left. \frac{\partial\ln
\varLambda(\lambda)}{\partial\lambda}\right|_{\lambda=0}
-\frac{99}{8}N.
\end{eqnarray}

\section{Nested algebraic Bethe ansatz}
\label{ABa}

In this section we show that the system (1) can be solved exactly by
using the nested algebraic Bethe ansatz method similar to that used
for the $spl(2|1)$ supersymmetric one
\cite{RamosMartins1996NPB96-474}.

The local vacuum state of the $i$th site is
chosen as $\left|0\right\rangle_i
=\left|3/2\right\rangle_i=(1, 0,0,0)^t$, where $t$
means the transport. The Lax operator on the site $i$
can be written into following matric from in the
auxiliary space
\begin{eqnarray}
L_{i}(\lambda) = \left(\begin{array}{llll}
^iA(\lambda)&^iB_1(\lambda)&
^iB_2(\lambda)&^iF(\lambda)\\
^iC^{1}(\lambda)&^iD^1_1(\lambda)&
^iD^1_2(\lambda)&^iB^{*1}(\lambda)\\
^iC^2(\lambda)&^iD^2_1(\lambda)&
^iD^2_2(\lambda)&^iB^{*2}(\lambda)\\
^iG(\lambda)&^iC_{1}^{*}(\lambda)&
^iC_{2}^{*}(\lambda)&^iV(\lambda)
\end{array}\right),
\end{eqnarray}
where $^iA$, $^iD^j_k$, $^iB_j$, $^iB^{*}_j$, $^iC$, $^iC_j$,
$^iC_j^{*}$, $^iG$, $^iV$ and $ ^iF$ are operators in the quantum
space $V_i$. The Lax operator $L_{i}$ acting on the local vacuum
state gives
\begin{eqnarray}
L_{i}(\lambda) \left|0\right\rangle =
\left(\begin{array}{rrrr} 1&^iB_1(\lambda)
\left|0\right\rangle&^iB_2(\lambda)
\left|0\right\rangle&^iF(\lambda)
\left|0\right\rangle\\
0&f(\lambda) \left|0\right\rangle & 0
&^iB^{*1}(\lambda)\left|0\right\rangle\\
0&0&f(\lambda) \left|0\right\rangle&
^iB^{*2}(\lambda)
\left|0\right\rangle\\
0&0&0&p(\lambda) \left|0\right\rangle
\end{array}\right),
\end{eqnarray}
where $ f(\lambda) =
2\lambda/(2\lambda+i)$.
The operators $^iB_1(\lambda)$ and $^iB_2(\lambda)$
are
\begin{eqnarray}
^iB_1(\lambda)=\left(\begin{array}{cccc}0&0&0&0\\
m(\lambda)&0&0&0\\0&0&0&0\\0&0&-s(\lambda)&0 \end{array}\right),~~~~~
^iB_2(\lambda)=\left(\begin{array}{cccc}0&0&0&0\\
0&0&0&0\\m(\lambda)&0&0&0\\0&s(\lambda)&0&0 \end{array}\right),
\end{eqnarray}
where $m(\lambda) = i/(2\lambda+i)$ and
$s(\lambda)=-2\lambda i/(2\lambda+3i)(2\lambda+i)$.
Thus the actions of $^iB_1$ and $^iB_2$ get a flip of
one and two spin quanta respectively. Obviously, the
whole state space can be obtained by these two
operators. This is quite different from the $SU(4)$
one in which $|-3/2\rangle_i$ can not be reached with
the flips made by $^iB_1$ and $^iB_2$. So that the
construction of the eigenstates might only need these
two operators in this model. This makes that
only once nesting is needed in the algebraic Bethe
ansatz of the model (1). In fact $^iF$ is also needed
in the construction of the eigenstates but plays an
assistant role and doesn't affect the number of
nestings.

The global vacuum state of the system (\ref{HSS}) is the production
of local vacuum states  $\left|0\right\rangle =
\left|0\right\rangle_1\otimes \left|0\right\rangle_2 \otimes \cdots
\otimes \left|0\right\rangle_N$. The monodromy matrix $T$ acting on
the global vacuum state gives
\begin{eqnarray}\label{monodromy-1}
&&T_N(\lambda) \left|0\right\rangle =
L_{N}(\lambda) L_{N-1}(\lambda) \cdots
L_{1}(\lambda) \left|0\right\rangle \nonumber \\
&&~~~~~~~~~~~~=\left(\begin{array}{cccc}
1, & \sum_i g^1_i(\lambda)
{~}^iB_1(\lambda), & \sum_{i} g^2_{i} (\lambda) {~}^iB_2(\lambda), &*\\
0, & f^N(\lambda), & 0, & \sum_{i} g^3_{i}(\lambda) {~}^iB^{*1}(\lambda)\\
0, & 0, & f^N(\lambda), & \sum_i g^4_i(\lambda) {~}^iB^{*2}(\lambda)\\
0, & 0, & 0, & p^N(\lambda) \end{array}\right)\left|0\right\rangle,
\end{eqnarray}
where $g_i^j$ are some coefficients, and $*$ represents the nonzero
element. The matric form of the monodromy matrix $T$ in the
auxiliary space is
\begin{eqnarray}\label{monodromy-a}
T(\lambda)= \left(\begin{array}{llll}
A(\lambda)&B_1(\lambda)&B_2(\lambda)&F(\lambda)\\
C^1(\lambda)&D^1_1(\lambda)&
{D^1_2}(\lambda)&B^{*1}(\lambda)\\
C^2(\lambda)&
D^2_1(\lambda)&
D^2_2(\lambda)&B^{*2}(\lambda)\\
G(\lambda)&C_1^*(\lambda)&C_2^*(\lambda)&V(\lambda)
\end{array}\right),
\end{eqnarray}
where$A$, $B_j$, $B^{*j}$, $G$, $C^j$, $C_j^{*}$, $D^j_k$, $F$ and
$V$ are operators in the Hilbert space $V_1\otimes V_2\otimes \cdots
\otimes V_N$. Acting these elements on the vacuum state, we obtain
\begin{eqnarray}
\label{oper-0} &&C^j(\lambda)\left|0\right\rangle
=0,~~~~
C^*_j(\lambda)\left|0\right\rangle=0,~~~~
G(\lambda)\left|0\right\rangle=0,~~~~
D^1_2(\lambda)\left|0\right\rangle=
D^2_1(\lambda)\left|0\right\rangle=0,\\
\label{oper-e} &&A(\lambda)\left|0\right\rangle
= \left|0\right\rangle, ~~~~
D^1_1(\lambda) \left|0\right\rangle =
D^2_2(\lambda)\left|0\right\rangle =
f^N(\lambda)\left|0\right\rangle, ~~~~
V(\lambda) \left|0\right\rangle
=  p^N(\lambda)\left|0\right\rangle.
\end{eqnarray}
The operators $C_j$, $C^*_j$, $G$, $D^2_1$ and $D^1_2$ acting on the
vacuum state $\left|0\right\rangle$ gives zero and the operators
$A$, $D^1_1$, $D^2_2$ and $V$ acting on the vacuum state
$\left|0\right\rangle$ give the eigenvalues. The elements $B_j$, $F$
and $B^{*j}$ acting on the vacuum state $\left|0\right\rangle$ will
generate other spin-flipped states, and thus can be regarded as the
generating operators of the multi-particle eigenstates of the
transfer matrix.

Now we turn to the eigenvalue problem of transfer
matrix $T$. From the definition (\ref{monodromy-a}),
the transfer matrix can be expressed in the form of
\begin{eqnarray}\label{transfer-t}
T(\lambda) = A(\lambda)+D^1_1(\lambda)+D^2_2(\lambda) +V(\lambda).
\end{eqnarray}
The eigenvalues of $T$ can be determined by the eigenvalues of
operators $B$, $D^1_1$, $D^1_2$ and $V$. These operators are shown
to be diagonalized in vacuum state $\left|0\right\rangle$ in Eq.
(\ref{oper-e}). Obviously,
$\left|\psi_0\right\rangle=\left|0\right\rangle$ is an eigenstate of
$t(\lambda)$, and the corresponding eigenvalue is
\begin{eqnarray}
\varLambda_0(\lambda) =\left[1+2f^N(\lambda)+p^N(\lambda)\right].
\end{eqnarray}
Here $\left|\psi_0\right\rangle$ is called zero-particle state. To
construct the other eigenstates of the transfer matrix, the
generating operators $B_j$, $B^{*j}$ and $F$ should be used. $B_j$
and $F$ are enough to  generate these states as shown bellow. Assume
that the eigenstates of the transfer matrix have the form
$\left|\psi\right\rangle = \psi(B_j, F)\left|0\right\rangle$. To go
further, we need need the commutation relations of
$A(\lambda)B_a(\mu)$, $D^a_b(\lambda)B_c(\mu)$, $V(\lambda)
B_a(\mu)$, $B(\lambda)F(\mu)$, $A^a_b(\lambda)F(\mu)$ and
$V(\lambda) F(\mu)$. From the Yang-Baxter relation (\ref{YBRRTT}) we
obtain \cite{RamosMartins1996NPB96-474}
\begin{eqnarray} \label{C-bB}
&& A(\lambda)B_a(\mu) = -\frac{\tilde m}{\tilde f} B_a(\lambda)
A(\mu) +\frac{1}{\tilde f} B_a(\mu) A(\lambda),
\\
\label{C-AB}
&& D^a_b(\lambda)B_c(\mu) = \frac{1}{f} B_e(\mu)
D^a_d(\lambda) r(\lambda-\mu)_{bc}^{ed} -\frac{m}{f} B_b(\lambda)
D^a_c(\mu) \nonumber\\ && ~~~~~~~~~~~~~~~~~~~ +\frac{s}{f}
\left[\frac{f}{p}B^{*a}A(\mu) + \frac{m}{p}F(\lambda) C^a(\mu) -
\frac{1}{p} F(\mu) C^a(\lambda)\right]\xi_{bc}.
\end{eqnarray}
Where $p(\lambda)
= 4\lambda(\lambda+i)/ (2\lambda+3i) (2\lambda+i)$, $\xi$ is a
vector of $2\otimes2$-dimension, $\xi = (0,1,-1,0)$, $r$ is the
matrix
\begin{eqnarray}
r(\lambda) = \left(\begin{array}{cccc}
1&0&0&0\\0&b(\lambda)&a(\lambda)&0\\
0&a(\lambda)&b(\lambda)&0\\0&0&0&1\end{array}
\right),
\end{eqnarray}
where $a(\lambda) = \lambda/(\lambda+i)$, $b(\lambda) =
i/(\lambda+i)$. Here the parameter $(\lambda-\mu)$ is omitted for short and the parameter $(\mu-\lambda)$ is also omitted by adding a tilde above the corresponding function. The commutation relation of $V(\lambda)B_a(\mu)$ is
\begin{eqnarray}
\label{C-dB} &&
V(\lambda)B_a(\mu) = \frac{f}{p}B_a(\mu)
V(\lambda) + \frac{m}{p}F(\mu)C^*_a(\lambda) -\frac{n}{p}
F(\lambda)C^*_a(\mu) -\frac{s}{p}\xi_{bc}B^{*b}
(\lambda)D^c_{a}(\mu),\nonumber\\
\end{eqnarray}
where $n(\lambda)= ic(4\lambda+3i)/(2\lambda+3i)(2\lambda+i)$.
\begin{eqnarray}
\label{C-bf} && A(\lambda)F(\mu) = \frac{1}{\tilde p}
F(\mu)A(\lambda) -\frac{\tilde n}{\tilde p}F(\lambda)A(\mu) +
\frac{\tilde s}{\tilde p} B_a(\lambda)B_b(\mu) \xi^{ab},
\\
\label{C-Af} &&
D^a_b(\lambda)F(\mu) = (1-\frac{m^2}{f^2})
F(\mu)D^a_b(\lambda) + \frac{m^2}{f^2}F(\lambda)D^a_b(\mu)
\nonumber\\ &&~~~~~~~~~~~~~~~~~~
+ \frac{m}{f}\left[B^{*a}(\lambda)
B_b(\mu) - B_b(\lambda) B^{*a}(\mu)\right].
\end{eqnarray}
The commutation relations of $D^a_a F$ are contained in them.
\begin{eqnarray}
\label{C-df} &&
V(\lambda)F(\mu) = \frac{1}{p}F(\mu)V(\lambda)
-\frac{n}{p}F(\lambda)V(\mu) -\frac{s}{p}\xi_{ab} B^{*a}(\lambda)
B^{*b}(\mu),
\end{eqnarray}
For $C_a$ and $C^*_a$ appear in the left hand of
(\ref{C-dB}), the commutation relation of them with
the operators $B_a$ and $F$ are also needed in the
discussion. For only one and two-particle cases are discussed in detail here, the
commutation rules used in this article are
\begin{eqnarray}
\label{C-C1B} && C^*_a(\lambda)B_b(\mu) = \frac{n}{p}
B_a(\mu)C^*_b(\lambda) + B_b(\mu)C^*_a(\lambda) \nonumber\\
&&~~~~~~~~~~ +\frac{s}{p} [A(\mu) V(\lambda)
-F(\mu)G(\lambda)]\xi_{ab} -\frac{n}{p} B_a(\lambda)C^*_b(\mu)-
\frac{s}{p} \xi_{cd} D^c_a(\lambda)D^d_{b}(\mu),\\
\label{C-CB}&&
C^a(\lambda)B_b(\mu) =B_b(\mu) C^a(\lambda) +
\frac{m}{f} [A(\mu) D_b^a(\lambda)-A(\lambda) D_b^a(\mu)],
\end{eqnarray}
\begin{eqnarray}
\label{C-BB} &&
B_a(\lambda) B_b(\mu) = B_c(\mu) B_d(\lambda)
r(\lambda-\mu)_{ab}^{cd} +\frac{s}{p}
[F(\lambda)A(\mu)-F(\mu)A(\lambda)]\xi_{ab},\\
\label{C-BB1} &&
B_a(\lambda) B^{*b}(\mu) = B^{*b}(\mu)
B_a(\lambda) +\frac{\tilde m}{\tilde f}
[F(\mu)D^b_a(\lambda)-F(\lambda)D^b_a(\mu)]\xi_{ab}.
\end{eqnarray}
Additionally, in the discussion of the three-particle wave
functions, the following commutation relations are needed,
\begin{eqnarray}
\label{C-Bf} &&
B_a(\lambda)F(\mu) = \frac{1}{\tilde f}F(\mu)
B_a(\lambda) - \frac{\tilde m}{\tilde f} F(\lambda)B_a(\mu).
\end{eqnarray}

From the commutation rules listed above, the eigenstates and
eigenvalues of the transfer matrices can be discussed. The one
particle state can be defined by $B_i$ operator acting on the vacuum
state as usual,
\begin{eqnarray}\label{state-1}
\left|\psi_1(\lambda_1)\right\rangle = B_a(\lambda_1) W^a \left|0\right\rangle,
\end{eqnarray}
where $W^a$ the coefficients of $B_a(\lambda_1)$. The transfer
matrix acting on the one particle state arrives
\begin{eqnarray}
&& t(\lambda)\left|\psi_1(\lambda_1)\right\rangle = [A(\lambda)
+D_{11}(\lambda)+D_{22}(\lambda) +V(\lambda)] B_a(\lambda_1) W^a
\left|0\right\rangle\nonumber\\
&& =\left\{ \frac{1}{f(\lambda_1-\lambda)}
+\frac{f(\lambda-\lambda_1)}{p(\lambda-\lambda_1)}
p^N(\lambda) +\frac{f^N(\lambda)}
{f(\lambda-\lambda_1)} [1+
a(\lambda-\lambda_1)]\right\}
\left|\psi_1(\lambda_1)\right\rangle
\nonumber\\
&&~~~-\left\{ \frac{m(\lambda_1-\lambda)}
{f(\lambda_1-\lambda)} + \frac{m(\lambda-\lambda_1)}
{f(\lambda-\lambda_1)} f^N(\lambda_1)\right\}
B_a(\lambda) W^a
\left|0\right\rangle\nonumber\\
&&~~~-\frac{s(\lambda-\lambda_1)}
{p(\lambda-\lambda_1)} \Big[
f_N(\lambda_1)-1\Big]  \xi_{ab} B^{*b}(\lambda) W^a
\left|0\right\rangle,
\end{eqnarray}
If the one particle state is a eigenstate of the transfer matrix,
terms including $B_a(\lambda)W^a\left|0\right\rangle$ and
$B_b^*(\lambda)\xi_{ab}W^a\left|0\right\rangle$ should be canceled
with each other. The condition that the unwanted terms cancel with
each other gives
\begin{eqnarray}\label{bae-1}
f^L(\lambda_1)-1=0.
\end{eqnarray}
The Eq. (\ref{bae-1}) is the Bethe ansatz equation. If the parameter
$\lambda_1 $ satisfies the Bethe ansatz equation (\ref{bae-1}), the
one particle state (\ref{state-1}) is an eigenstate of the system.
The eigenvalue $\varLambda_1(\lambda, \lambda_1)$ of the transfer
matrix at one-particle state is
\begin{eqnarray}\label{eigen-1}
\varLambda_1(\lambda,
\lambda_1) = \frac{1}{f(\lambda_1-\lambda)}
+\frac{f(\lambda-\lambda_1)}{p(\lambda-\lambda_1)}
p^N(\lambda) +\frac{f^N(\lambda)}
{f(\lambda-\lambda_1)} [1+
a(\lambda-\lambda_1)].
\end{eqnarray}

To generate the two particle states, operator
$F$ is needed additionally. The two-particle
eigenstate is assumed to be
\begin{eqnarray}\label{state-2}
&& \left|\psi_2(\lambda_1,\lambda_2)\right\rangle =
B_{a_1}(\lambda_1) B_{a_2}(\lambda_2) W^{a_2a_1}\left|0\right\rangle
+ h(\lambda_1, \lambda_2) F (\lambda_1) \xi_{a_2a_1} W^{a_2a_1}
\left|0\right\rangle,
\end{eqnarray}
where $h(\lambda_1,\lambda_2)$ is a undetermined function. Acting
the transfer matrix $t_N(\lambda)$  on the assumed state
(\ref{state-2}), we obtain
\begin{eqnarray}
&& t_N(\lambda) \left|\psi_2(\lambda_1, \lambda_2)\right\rangle =
\left[A(\lambda) +D^1_1(\lambda) +D^2_2(\lambda)
+V(\lambda)\right] \left|\psi_2(\lambda_1,\lambda_2)\right\rangle
\nonumber \\
&&
~~~~~~~~~~~~~~~~~~~~~~~~ = \left|\psi_2^0\right\rangle +
 \left|\psi_2^1\right\rangle +\left|\psi_2^2\right\rangle
+\left|\psi_2^3\right\rangle+ \left|\psi_2^4\right\rangle+
\left|\psi_2^5\right\rangle +\left|\psi_2^6\right\rangle
+\left|\psi_2^7\right\rangle.
\end{eqnarray}
Here, $\left|\psi_2^0\right\rangle$ denotes the eigenstate which
including the operators $B_{a_1} (\lambda_1) B_{a_2} (\lambda_2)$
and $F (\lambda_1)$. $\left|\psi_2^1\right\rangle$,
$\left|\psi_2^2\right\rangle$, $\left|\psi_2^3\right\rangle$,
$\left|\psi_2^4\right\rangle$, $\left|\psi_2^5\right\rangle$,
$\left|\psi_2^6\right\rangle$ and $\left|\psi_2^7\right\rangle$
denote the unwanted terms including $B^*_{a_1}(\lambda)
B^*_{a_2}(\lambda_1)$, $B_{m}(\lambda)B^*_{m}(\lambda_1)$,
$B_{a_1}(\lambda) B_{a_2}(\lambda_2)$, $B^*_{m}(\lambda)
B_{a_2}(\lambda_2)$, $B_{a_1}(\lambda) B_{a_2}(\lambda_1)$,
$B^{*m}(\lambda) B_{c}(\lambda_1)$ and $F(\lambda)$ respectively.
The unwanted terms should be canceled with each other, which gives
the form of the function $h$ and the Bethe ansatz equations. The
unwanted terms $\left|\psi^1_2\right\rangle$ and
$\left|\psi_2^2\right\rangle$ can be explicitly expressed as
\begin{eqnarray}
&&\left|\psi_2^1\right\rangle =
\frac{s(\lambda-\lambda_1)} {p(\lambda-\lambda_1)}
\left[ h(\lambda_1,\lambda_2)-\frac{s(\lambda_1-\lambda_2)}
{p(\lambda_1-\lambda_2)}
\right] \xi_{b_1b_2}
W^{b_1b_2}\xi_{a_1a_2}
B^{*a_1}(\lambda)B^{*a_2}(\lambda_2) \left|0\right\rangle,\nonumber\\ \\
&&\left|\psi_2^2\right\rangle = \frac{m(\lambda-\lambda_1)}
{f(\lambda-\lambda_1)}\hspace{-3.5pt} \left[ h(\lambda_1,\lambda_2)
-\frac{s(\lambda_1-\lambda_2)} {p(\lambda_1-\lambda_2)}
\right]\xi_{a_1a_2} W^{a_1a_2} B_{m}(\lambda)B^*_{m}(\lambda_2)
\left|0\right\rangle.
\end{eqnarray}
If the function $h$ satisfies
\begin{eqnarray}\label{h}
h(\lambda_1,\lambda_2) = \frac{s(\lambda_1-\lambda_2)}
{p(\lambda_1-\lambda_2)},
\end{eqnarray}
the terms $\left|\psi^1_2\right\rangle$ and
$\left|\psi_2^2\right\rangle$ are canceled with each other. This
constraint determines the values of $h$ introduced in the
two-particle eigenstate (\ref{state-2}). The unwanted terms
$\left|\psi_2^3\right\rangle$ and $\left|\psi_2^4\right\rangle$ are
\begin{eqnarray}
&&\left|\psi_2^3\right\rangle =
\left[-\frac{m(\lambda_1-\lambda)}
{f(\lambda_1-\lambda)f(\lambda_2-\lambda_1)}
W^{a_1a_2} \right.\nonumber\\
&&~~~~~~~~~\left.-\frac{m(\lambda-\lambda_1)
f^N(\lambda_1)}{f(\lambda-\lambda_1)
f(\lambda_1-\lambda_2)}
r(\lambda_1-\lambda_2)_{b_1b_2}^{a_2a_1}
W^{b_1b_2}\right] B_{a_1}(\lambda)
B_{a_2} (\lambda_2) \left|0\right\rangle,\\
&&\left|\psi_2^4\right\rangle = \left[-
\frac{s(\lambda-\lambda_1)f^N(\lambda_1)}
{f(\lambda-\lambda_1)f(\lambda_1-\lambda_2)} \xi_{ma_1}
r(\lambda_1-\lambda_2)_{b_1b_2}^{a_2a_1}
W^{b_1b_2}\right.\nonumber\\
&&~~~~~~~~~\left. +\frac{s(\lambda-\lambda_1)}
{p(\lambda-\lambda_1)f(\lambda_2-\lambda_1)} \xi_{ma_1}
W^{a_1a_2}\right] B^*_{m}(\lambda)B_{a_2}(\lambda_2)
\left|0\right\rangle.
\end{eqnarray}
After some tedious calculations, we find that if the parameter
$\lambda_1$ and $\lambda_2$ satisfy the following Bethe ansatz
equations
\begin{eqnarray}\label{bae-2-1}
\frac{1}{f^N(\lambda_1)} \frac{f(\lambda_1-\lambda_2)}
{f(\lambda_2-\lambda_1)} W^{a_1a_2} = r(\lambda_1-\lambda_2)_{b_1b_2}^{a_2a_1}
W^{b_1b_2}.,
\end{eqnarray}
the terms $\left|\psi^3_2\right\rangle$ and
$\left|\psi_2^4\right\rangle$ are also canceled with each other. The
unwanted term $\left|\psi_2^5\right\rangle$ is
\begin{eqnarray}
&&\left|\psi_2^5\right\rangle = \left\{
h(\lambda_1,\lambda_2) \frac{s(\lambda_1-\lambda)}
{p(\lambda_1-\lambda)} \xi_{b_2b_1}\xi^{a_1a_2}
W^{b_1b_2} \right. + \frac{m(\lambda_1-\lambda)
m(\lambda_2-\lambda_1)} {f(\lambda_1-\lambda)
f(\lambda_2-\lambda_1)}
W^{a_1a_2}\nonumber\\
&& ~~~~~~~~~ -\frac{m(\lambda_2-\lambda)}
{f(\lambda_2-\lambda) f(\lambda_1-\lambda)}
r(\lambda_1-\lambda)_{b_1b_2}^{a_1a_2}
W^{b_1b_2}\nonumber\\
&& ~~~~~~~~~ -\frac{m(\lambda-\lambda_2)}
{f(\lambda-\lambda_1)f(\lambda -\lambda_2)}
r(\lambda_1-\lambda)_{b_1b_2}^{a_1a_2}
r(\lambda-\lambda_1)_{c_1c_2}^{b_1b_2} f^N(\lambda_2)
W^{c_2c_1}\nonumber \\
&& ~~~~~~~~~
\left.+\frac{m(\lambda-\lambda_1)m(\lambda_1-\lambda_2)}
{f(\lambda-\lambda_1)f(\lambda_1-\lambda_2)}
f^N(\lambda_2) W^{a_2a_1} \right\}
B_{a_1}(\lambda_1) B_{a_2}(\lambda),
\end{eqnarray}
We find that if the parameter $\lambda_1$ and $\lambda_2$ satisfy
Eq. (\ref{h}) and
\begin{eqnarray}\label{bae-2-2}
\frac{1}{f^N(\lambda_2)}
\frac{f(\lambda_2-\lambda_1)} {f(\lambda_1-\lambda_2)} W^{a_1a_2} =
r(\lambda_2-\lambda_1)^{a_2a_1}_{b_1b_2}
W^{b_1b_2}.
\end{eqnarray}
The unwanted term $\left|\psi^5_2\right\rangle$ is zero. The Eqs.
(\ref{bae-2-1}) and (\ref{bae-2-2}) are the two-particle Bethe
ansatz equations, which can be written into a uniform form
\begin{eqnarray}\label{bae-2}
\frac{1}{f^N(\lambda_i)}
\frac{f(\lambda_i-\lambda_j)} {f(\lambda_j-\lambda_i)}
W^{a_1a_2} = r(\lambda_i-\lambda_j)_{b_1b_2}
^{a_2a_1} W^{b_1b_2}, \hspace{10pt}
(i,j=1,2,\hspace{5pt}i\neq j).
\end{eqnarray}
The function $W^{a_1a_2}$ can be determined by the nested algebraic
Bethe ansatz. Together with Eq. (\ref{h}), the following unwanted
terms $\left|\psi_2^6\right\rangle$ and $\left|\psi_2^7\right\rangle$
can be canceled,
\begin{eqnarray}
&&\left|\psi_2^6\right\rangle =\left\{\Big[ \frac{s(\lambda
-\lambda_1)}{p(\lambda -\lambda_1)} \frac{m(\lambda_1
-\lambda_2)}{f(\lambda_1 -\lambda_2)} -\frac{f(\lambda
-\lambda_1)}{p(\lambda -\lambda_1)} \frac{s(\lambda
-\lambda_2)}{p(\lambda -\lambda_2)} \Big] f^N(\lambda_2)
\xi_{ma_2}W^{ca_2}
\right. \nonumber\\
&&\left.\Big[ -\frac{s(\lambda -\lambda_1)}{p(\lambda -\lambda_1)}
\frac{m(\lambda_2 -\lambda_1)}{f(\lambda_2 -\lambda_1)}
\xi_{ma_1}W^{a_1c} + \frac{1}{f(\lambda -\lambda_1)} \frac{s(\lambda
-\lambda_2)}{p(\lambda -\lambda_2)} \xi_{b_2a_2} r(\lambda
-\lambda_1)^{cb_2}_{ma_1}\right\} \nonumber\\
&& \times W^{a_1a_2}\Big] B^{*m}(\lambda) B_c(\lambda_1) \left|0\right\rangle +
h(\lambda_1,\lambda_2) \frac{m(\lambda -\lambda_1)}{f(\lambda
-\lambda_1)} \xi_{a_2a_1}
W^{a_1a_2} B^{*m}(\lambda) B_{m}(\lambda_1) \left|0\right\rangle,
\nonumber\\ \\
&&\left|\psi_2^7\right\rangle =\frac{n(\lambda-\lambda_1)}{p(\lambda-\lambda_1)}
\Big[ h(\lambda_1, \lambda_2)- \frac{s(\lambda_1-\lambda_2)}{p(\lambda_1-\lambda_2)}\Big]
p^N(\lambda_1) F(\lambda) \xi_{a_1a_2} W^{a_1a_2}\left|0\right\rangle
\nonumber\\
&&+ \Big[ \frac{1}{f(\lambda_1-\lambda)}
\frac{m(\lambda_2-\lambda)}{f(\lambda_2-\lambda)}
\frac{s(\lambda_1-\lambda)}{p(\lambda_1-\lambda)}+h(\lambda_1,
\lambda_2) \frac{n(\lambda_1 -\lambda)}{p(\lambda_1 -\lambda)}\Big]
\xi_{a_1a_2}W^{a_1a_2}\nonumber\\
&&\times F(\lambda) \left|0\right\rangle +\Big[ \frac{n(\lambda -\lambda_1)}{p(\lambda -\lambda_1)}
\frac{s(\lambda_1 -\lambda_2)}{p(\lambda_1 -\lambda_2)}
-\frac{m(\lambda -\lambda_1)}{p(\lambda -\lambda_1)} \frac{s(\lambda
-\lambda_2)}{p(\lambda -\lambda_2)} \Big] f^N(\lambda_1)
f^N(\lambda_2) \xi_{a_1a_2}\nonumber\\
&&\times W^{a_1a_2} F(\lambda) \left|0\right\rangle +2h(\lambda_1,
\lambda_2) \frac{m^2(\lambda-\lambda_1)} {f^2(\lambda-\lambda_1)}
f^N(\lambda_1) \xi_{a_2a_1} W^{a_1a_2} F(\lambda)
\left|0\right\rangle
\nonumber\\
&&+\frac{1}{f(\lambda -\lambda_1)} \frac{m(\lambda
-\lambda_1)}{p(\lambda -\lambda_1)} \frac{s(\lambda
-\lambda_2)}{p(\lambda -\lambda_2)} r(\lambda -\lambda_1) ^{b_1b_2}
_{b_1a_1} f^N(\lambda_1)\xi_{b_2a_2} W^{a_1a_2} F(\lambda)
\left|0\right\rangle\nonumber\\
&&+\frac{s(\lambda -\lambda_1)}{f(\lambda -\lambda_1)}
\frac{m(\lambda -\lambda_1)}{p(\lambda -\lambda_1)}
\frac{m(\lambda_1 -\lambda_2)}{f(\lambda_1 -\lambda_2)} \xi_{a_2a_1}
[f^N(\lambda_1) -f^N(\lambda_2)] W^{a_1a_2}
F(\lambda)
\left|0\right\rangle\nonumber\\
&&+\frac{1}{f(\lambda -\lambda_1)} \frac{m(\lambda
-\lambda_2)}{p(\lambda -\lambda_2)} \frac{s(\lambda_1
-\lambda)}{p(\lambda_1 -\lambda)} r(\lambda -\lambda_1) ^{b_1b_2}
_{a_2a_1} f^N(\lambda_2)\xi_{b_1b_2} W^{a_1a_2} F(\lambda)
\left|0\right\rangle.
\end{eqnarray}

With the above results, the explicit form of the two-particle
eigenstate of the system is
\begin{eqnarray}
&&\left|\Psi_2^0\right\rangle= \left\{ \frac{1}
{f(\lambda_1-\lambda) f(\lambda_2-\lambda)} +
\frac{f(\lambda-\lambda_1) f(\lambda-\lambda_2)}
{p(\lambda_1-\lambda) p(\lambda_2-\lambda)}
p^N(\lambda) \right.\nonumber\\
&& ~~~~~~ \left.+\frac{f^N(\lambda)}
{f(\lambda-\lambda_1)f(\lambda-\lambda_2)}
r^{c_2d_1}_{a_1b_2}(\lambda-\lambda_1)
r^{b_2d_2}_{a_2c_2}(\lambda-\lambda_2) \right\}
B_{a_1}(\lambda_1) B_{a_2}(\lambda_2) W^{a_2a_1}
\left|0\right\rangle \nonumber\\
&& ~~~~~~
 +\left\{h(\lambda_1,\lambda_2)
\left[\frac{1} {p(\lambda_1 -\lambda)} +
\frac{p^N(\lambda)} {p(\lambda-\lambda_1)} +
\left(1-\frac{m^2(\lambda -\lambda_1)} {f^2(\lambda
-\lambda_1)}\right) 2f^N(\lambda) \right]\right.
\xi_{a_2a_1}\nonumber\\
&&~~~~~~
+\left[ \frac{1}{f(\lambda_1-\lambda)}
\frac{m(\lambda_2-\lambda)} {f(\lambda_2-\lambda)}
\frac{s(\lambda_1-\lambda)}
{p(\lambda_1-\lambda)} +
\frac{s(\lambda-\lambda_2)}
{p(\lambda-\lambda_2)} \frac{m(\lambda-\lambda_1)}
{f(\lambda-\lambda_1)}
p^N(\lambda)\right.\nonumber\\
&&~~~~~~
+\left.\frac{s(\lambda-\lambda_1)}
{f(\lambda-\lambda_1) p(\lambda-\lambda_1)}
\frac{m(\lambda-\lambda_2)}
{f(\lambda-\lambda_2)}\left[ f^N(\lambda_2) -
f^N(\lambda) \right] \right] \xi_{a_2a_1} \nonumber\\
&&~~~~~~
+\frac{1}{f(\lambda-\lambda_1)} \left[
\frac{m(\lambda-\lambda_2)}{f(\lambda-\lambda_2)}
\frac{s(\lambda_1-\lambda)}{p(\lambda_1-\lambda)}
r^{a_2a_1}_{b_1b_2}(\lambda-\lambda_1)
\xi_{b_1b_2}f^N(\lambda_2)\right.\nonumber\\
&&~~~~~~
\left.\left. +\frac{s(\lambda-\lambda_2)}{p(\lambda-\lambda_2)}
\frac{m(\lambda_1-\lambda)}{p(\lambda_1-\lambda)}
r^{b_1a_1}_{b_1b_2}(\lambda-\lambda_1) \xi_{b_2a_2}
f^N(\lambda)\right]\right\}  F(\lambda_1) W^{a_2a_1}
\left|0\right\rangle.
\end{eqnarray}
Considering Eqs. (\ref{h}) and (\ref{bae-2}), the eigen-equation of
the transfer matrix becomes
\begin{eqnarray}
&&t(\lambda)\left|\psi_2(\lambda_1,\lambda_2)\right\rangle = \prod_{i=1}^2\frac{1} {f(\lambda_i-\lambda)}
+p^N(\lambda) \prod_{i=1}^2
\frac{f(\lambda-\lambda_i)}{p(\lambda-\lambda_i)}
+\prod_{i=1}^2
\frac{f^N(\lambda)}{f(\lambda-\lambda_i)}
\nonumber\\
&&\times r(\lambda-\lambda_1)_{c_2d_1}^{a_1b_2}
r(\lambda-\lambda_2)_{b_2d_2}^{a_2c_2}
W^{d_1d_2}
[B_{a_1}(\lambda_1) B_{a_2}(\lambda_2) +
h(\lambda_1,\lambda_2) F (\lambda_1) \xi_{a_2a_1}
]\left|0\right\rangle.
\end{eqnarray}
The eigen-value of two-particle state is
\begin{eqnarray}
&& \varLambda_2(\lambda, \{\lambda_1, \lambda_2\}) =
\prod_{i=1}^2 \frac{1} {f(\lambda_i-\lambda)}
+ p^N(\lambda) \prod_{i=1}^2
\frac{f(\lambda-\lambda_i)} {p(\lambda-\lambda_i)}
\nonumber\\
&&~~~~~~~~~~~~~~~~~~~~~~~+ f^N(\lambda)
\prod_{i=1}^2\frac{1}{f(\lambda-\lambda_i)} \varLambda^{(1)}_2
(\lambda, \{\lambda_1, \lambda_2\}),
\end{eqnarray}
where $\varLambda^{(1)}_2$ is the eigenvalue of $t^{(1)}(\lambda,
\{\lambda_1,\lambda_2\})$.

From the above discussions, we see that the construction of
eigenstates of the $SO(5)$-invariant quantum spin chain is quit
different from that of the $SU(4)$-invariant one. The spin-flipped
operators are very complicated. The symmetry analysis of the states
is helpful to construct the eigenstates
\cite{RamosMartins1996NPB96-474}.  Now, we construct the
many-particle eigenstates. We define $\vec \psi_n$ as the creation
operators of the $n$-particle state for convenience
\begin{eqnarray}
\left|\psi_n(\{\lambda_i\})\right\rangle = \vec \psi_n \vec W \left|0\right\rangle,
\end{eqnarray}
where $\vec W = (W^1, W^2,\cdots,W^n)^T$ are some vectors.
Obviously, $\vec \psi_0 = 1$ for $\left|\psi_0\right\rangle
=\left|0\right\rangle$. From Eqs. (\ref{state-1}) and
(\ref{state-2}), we know the creation operators of one and
two-particle eigenstates are
\begin{eqnarray}
\vec\psi_1= \vec B(\lambda_1), ~~~
\vec \psi_2(\lambda_1,\lambda_2) = \vec B(\lambda_1)  \otimes \vec
B(\lambda_2) + h(\lambda_1, \lambda_2) F(\lambda_1)  \vec \xi.
\end{eqnarray}
From the commutation relations (\ref{C-BB}), the relation of $\vec
\psi(\lambda_1, \lambda_2)$ and $\vec \psi(\lambda_2,\lambda_1)$ is
\begin{eqnarray}
\vec \psi_2 (\lambda_2, \lambda_1) = \vec \psi_2 (\lambda_1,
\lambda_2)  r(\lambda_2-\lambda_1),~~~ \vec \psi_2
(\lambda_1, \lambda_2) = \vec \psi_2 (\lambda_2, \lambda_1)
r(\lambda_1-\lambda_2).
\end{eqnarray}
After a more detailed analysis, we find that the three-particle
states should have the same symmetry. Assume the three-particle
eigenstates as
\begin{eqnarray}
&&\vec\psi_3(\lambda_1,\lambda_2,\lambda_3) = \vec B(\lambda_1)
\otimes \vec \psi_2(\lambda_2, \lambda_3) - h(\lambda_1,\lambda_2)
F(\lambda_1) \vec \xi \otimes \vec \psi_1(\lambda_3)
\frac{1}{f(\lambda_3-\lambda_2)}\nonumber\\
&& ~~~~~~~~~~~~~~~~~~~~~
- h(\lambda_1,\lambda_3)\hat
F(\lambda_1) \vec \xi \otimes \vec \psi_1(\lambda_2) \frac{1}{f(\lambda_2-\lambda_3)}
r_{23}(\lambda_2-\lambda_3),
\end{eqnarray}
which satisfies
\begin{eqnarray}
&&\left|\psi_3(\lambda_2,\lambda_1,\lambda_3)\right\rangle =
\left|\psi_3(\lambda_1, \lambda_2,\lambda_3)\right\rangle
r_{12}(\lambda_1-\lambda_2), \nonumber \\
&&\left|\psi_3(\lambda_1,\lambda_3,\lambda_2)\right\rangle =
\left|\psi_3(\lambda_1,\lambda_2,\lambda_3)\right\rangle
r_{23}(\lambda_2-\lambda_3).
\end{eqnarray}
Borrowed the ideas in \cite{RamosMartins1996NPB96-474}, the
generator of multi-particle state is assumed as
\begin{eqnarray}
&&\vec \psi_{M_1}(\lambda_1, \lambda_2, \cdots, \lambda_{M_1}) =
\vec B(\lambda_1)\otimes\vec \psi_{{M_1}-1} (\lambda_2, \lambda_3,
\cdots, \lambda_{M_1})\nonumber\\
&&~~~~~~~~~~~~~~~~~
-  F(\lambda_1)
\vec \xi \otimes \sum_{j=2}^{M_1} h(\lambda_1-\lambda_j) \prod_{k =
2, k\neq j}^n \frac{1}{f(\lambda_k-\lambda_j)}\nonumber\\
&&~~~~~~~~~~~~~~~~~ \times \vec \psi_{{M_1}-2} (\lambda_2, \cdots,
\lambda_{j-1}, \cdots, \lambda_{j+1}, \lambda_{M_1})
\prod_{k=2}^{j-1} r_{k,k+1}(\lambda_k-\lambda_j),
\end{eqnarray}
which satisfies
\begin{eqnarray}
\vec \psi_{M_1}(\lambda_1, \lambda_2, \cdots, \lambda_{M_1}) =  \vec
\psi_{M_1}(\lambda_1, \cdots,\lambda_{j+1}, \lambda_{j}, \cdots,
\lambda_{M_1}) r_{j,j+1} (\lambda_j-\lambda_{j+1}).
\end{eqnarray}
By using the same process as that for the two-particle case, we
obtain the following Bethe ansatz equations
\begin{eqnarray}\label{BAe-1}
&&\frac{1}{f^N(\lambda_j)} W^{a_1\cdots a_{M_1}} = \prod_{i\neq
j}^{M_1} \frac{f(\lambda_i-\lambda_j)}{f(\lambda_j-\lambda_i)}
\nonumber\\
&&~~~~~~~~~~~~~~~~~~~~~~~~~~
\times\varLambda^{(1)}_{M_1-1}(\lambda_j, \{\lambda_{j+1},
\cdots\lambda_{M_1},\lambda_{1}, \cdots\lambda_{j-1}\})
W^{d_1\cdots d_{M_1}},\\
&&\label{Egv-1-1}
\varLambda_{M_1} (\lambda, \{\lambda_i\}) =
\prod_{k=1}^{M_1}  \frac{1}{f(\lambda_k-\lambda)}
\varLambda^{(1)}_{M_1}(\lambda, \{\lambda_i\}),
\end{eqnarray}
where $\varLambda^{(1)}_{M_1}(\lambda, \{\lambda_i\})$ is the
eigenvalue of a series production of $r$ matrices
\begin{eqnarray}\label{Egv-1-2}
&&\hspace{-12pt} r(\lambda-\lambda_1)_{md_1}^{a_1b_1}
r(\lambda-\lambda_2) _{b_1d_2}^{a_2b_2} \cdots r(\lambda-\lambda_n)
_{b_{n-1}d_n}^{a_nm} W^{d_1\cdots d_n} =
\varLambda^{(1)}_{M_1}(\lambda, \{\lambda_i\}) W^{a_n\cdots a_1}.
\end{eqnarray}

Now, we diagonalize the eigen-equation (\ref{Egv-1-2}). One can
easily check that the nested $R$ matrix satisfies the Yang-Baxter
equation (\ref{YBERRR}). In fact, it is the $6$-vertex $R$-matrix in
space $V^{(1)} \otimes V^{(1)} $ with $V^{(1)}$ a two dimensional
space. The corresponding Lax operator is
\begin{eqnarray}
L^{(1)}_{12}(\lambda) = P_{12} r_{12} (\lambda).
\end{eqnarray}
The nested monodromy matrix is defined as
\begin{eqnarray}\label{monodromyn}
{T}^{(1)}_{M_1}(\lambda,  \{\lambda_j\}) =
{L}^{(1)}_{0,{M_1}}(\lambda-\lambda_1)
{L}^{(1)}_{0,{M_1}-1}(\lambda-\lambda_2) \cdots
{L}^{(1)}_{0,1}(\lambda-\lambda_{M_1}),
\end{eqnarray}
which satisfies the Yang-Baxter relation (\ref{YBRRTT}). The nested
transfer matrix is
\begin{eqnarray}\label{transfern}
t^{(1)}_{M_1}(\lambda) = \mathrm{tr}_0 T^{(1)}_{M_1}(\lambda).
\end{eqnarray}
The transfer matrices with different spectral parameters commutate
with each other
\begin{eqnarray}
[t^{(1)}_{M_1}(\lambda),t^{(1)}_{M_1}(\mu)]=0.
\end{eqnarray}
Using the standard Bethe ansatz method for the six-vertex model, we
obtain the eigenvalue of nested transfer matrix $t^{(1)}_{M_1}$ as
\begin{eqnarray}\label{egvn}
\varLambda^{(1)}_{M_2}(\lambda, \{\mu_j\}) = \prod_{i=1}^{M_2}
\frac{1} {a(\mu_i-\lambda)} + \prod_{i=1}^{M_1} a(\lambda-\lambda_i)
\prod_{i=1}^{M_2} \frac{1}{a(\lambda-\mu_i)},
\end{eqnarray}
where the parameter $\mu_i$ should satisfy the following Bethe
ansatz equation
\begin{eqnarray}\label{baen}
\prod_{i=1, \neq j}^m\frac{a(\mu_j-\lambda_i)}{a(\mu_i-\lambda_j)} =
\prod_{i=1}^na(\mu_j-\lambda_i).
\end{eqnarray}

Substituting Eq. (\ref{egvn}) into (\ref{BAe-1}) and
(\ref{Egv-1-1}), we obtain the eigenvalues of $T_N(\lambda)$ as
\begin{eqnarray}\label{EGV-1}
&&\varLambda_{M_1M_2}(\lambda,
\{\lambda_i\},  \{\mu_j\}) =\prod_{i=1}^{M_1}
\frac{1}{f(\lambda_i-\lambda)}+p^N(\lambda) \prod_{i=1}^{M_1}
\frac{f(\lambda-\lambda_i)} {p(\lambda-\lambda_i)} \nonumber\\
&&~+f^N(\lambda) \prod_{i=1}^{M_1} \frac{1} {f(\lambda-\lambda_i)}
\left[ \prod_{i=k}^{M_1} a(\lambda-\lambda_k) \prod_{j=1}^{M_2}
\frac{1}{a(\lambda-\mu_j)} +\prod_{j=1} ^{M_2} \frac{1}{a(\mu_j -
\lambda)}\right]\left|\psi_{M_1}\right\rangle.
\end{eqnarray}
The Bethe ansatz equations are Eq. (\ref{baen}) and
\begin{eqnarray}
\label{BAE0-1} && \frac{1}{f^L(\lambda_j)} = \prod_{i\neq j}^{M_1}
\frac{ f(\lambda_i -\lambda_j)} {f(\lambda_j -\lambda_i)}
\prod_{k=1}^{M_1} \frac{1} {a(\mu_k -\lambda_j)}.
\end{eqnarray}
Putting $\lambda_i \to \lambda_i - i/4$, $\mu_j
\to \mu_j + {3i}/{4} $,  the Bethe ansatz
equations (\ref{baen}) and (\ref{BAE0-1}) can be written as
\begin{eqnarray}
\label{BAE1}
&& \left[\frac{\lambda_j+i/4}{\lambda_j-i/4}\right]^N
\prod_{i=1}^{{M_2}} \frac{\lambda_j -\mu_i+i/2}{\lambda_j
-\mu_i-i/2} = -\prod_{i=1}^{M_1}
\frac{\lambda_j-\lambda_i+i/2}{\lambda_j
-\lambda_i-i/2},
\nonumber\\
&&\prod_{i=1}^{M_1} \frac{\mu_j-\lambda_i+i/2} {\mu_j
-\lambda_i-i/2} = -\prod_{i=1} ^{M_2}
\frac{\mu_j-\mu_i+i} {\mu_j-\mu_i-i}.
\end{eqnarray}

From the relation (\ref{EGV-1}), we obtain the eigenvalue $E$ of the
Hamiltonian $H$ (\ref{HSS}) as
\begin{eqnarray} \label{energy-H-1}
E = -\frac{9J}{4} \left( \left. \frac{\partial \ln T(\lambda)}
{\partial \lambda} \right|_{ \lambda=0} +\frac{11}{2}N \right)=
-\frac{9J\pi}{2}  \sum_{j=1}^{M_1} a_{1/4} (\lambda_j)
-\frac{99}{8}JN,
\end{eqnarray}
where $a_t(x) = t/[\pi(x^2+t^2)]$ and $\lambda_j$ should satisfy the
Bethe ansatz equations (\ref{BAE1}). The momentum $P$ is
\cite{Takhatajan1982, Sutherland2004}
\begin{eqnarray}\label{momentum-P-1}
P = \sum_{j=1}^{M_1} k_j \mod 2\pi,
\end{eqnarray}
where $k_j$ are parameterized by $e^{ik_j} =
(\lambda_j-i/4)/(\lambda_j+i/4)$.

The basis of Hilbert space $V_i$ are
$\left|3/2\right\rangle_i$,
$\left|1/2\right\rangle_i$,
$-\left|1/2\right\rangle_i$ and
$-\left|3/2\right\rangle_i$. Thus the vacuum state is
a ferromagnetic state $\left|0\right\rangle =
\left|3/2\right\rangle_1\otimes
\left|3/2\right\rangle_2 \otimes\cdots\otimes
\left|3/2\right\rangle_N$. The $M_1 +M_2$ is the
total number of flipped spins from this ferromagnetic
state, so the total spin along the $z$-component and
magnetization are
\begin{eqnarray}\label{Sz}
&&S^z = 3N/2 - (M_1+M_2),\\
&&\label{mag}
\mathfrak{m} = 3/2 - (M_1 + M_2)/N =
3/2 - n_{\lambda} -n_{\mu}.
\end{eqnarray}
From Eq. (\ref{energy-H-1}), if $J<0$, the $\lambda$ will lead to
addition to the energy. Therefore, the ground state of this case is
the ferromagnetic state $\left|0\right\rangle$. While if $J>0$, the
$\lambda$ will contribute a negative value to the energy, so the
ground state might be a anti-ferromagnetic state. In order to obtain
the ground state configuration of the system, we first consider the
thermodynamics. The ground state could be obtained by putting the
temperature to zero.

\section{Thermodynamics}
\label{TD}

Now, we solve the Bethe ansatz equations (\ref{BAE1}).
In the thermodynamic limit, both the total particle
number and the system size tend to infinity while the
ratio $N/L$ keeps a non-zero constant. The Bethe
ansatz equations can have the complex solutions, i.e.,
strings. Checking of the Bethe ansatz equations
(\ref{BAE1}) in detail, we find that the string hypothesis
of the Bethe ansatz
equations (\ref{BAE1}) is
\begin{eqnarray}
&& \lambda^{n}_{\beta,j} = \lambda^{n}_\beta +
\frac{i}{4}(n+1-2j), \; j=1,\cdots,n, \nonumber \\
&& \mu^{m}_{\nu,j} = \mu^{m}_\nu + \frac{i}{2}(n+1-2j),\;
j=1,\cdots,m, \label{string_h}
\end{eqnarray}
where $\lambda^{n}_\beta$ and $\mu^{m}_\nu$ are the real parts
of the $n$-string of $\lambda$ and the $m$-string of $\mu$,
respectively. From Eq. (\ref{EGV-1}), we see that the contributions
of $n$-string of $\lambda$ to the energy is
\begin{eqnarray}\label{eGVH1-1}
e_n(\lambda^{m}_\beta) = - \frac{9}{2}J \pi a_{n/4}
(\lambda^{n}_z).
\end{eqnarray}
The energy $E$ and the momentum $P$ are
\begin{eqnarray}
\label{energy-H-2} E &=&  \sum_{n,z} e_n(\lambda_z^{n}) -
\frac{99}{8}JN = -\frac{9}{2} J\pi \sum_{n,z}
a_{n/4}(\lambda_z^{n}) - \frac{99}{8}JN, \\ P &=&  \sum_{n,z} \pi-
\theta_{n/4}(\lambda^{n}_j) \mod 2\pi,
\end{eqnarray}
where $(x-it)/(x+it) = -e^{i\theta_t(x)}$ and $\theta_{t}(x) =
2\arctan(x/t)$.

Substituting the string solutions into the Bethe ansatz equations
(\ref{BAE1}), we have
\begin{eqnarray}
&& e^{-i N\theta_{\frac{n}{4}}(\lambda^n_z)} \prod_{m,y}
e^{-i\mathcal{B}^4_{m,n}(\lambda^n_z-\mu^m_y)}= (-)
^{\delta_{\lambda^n}}
\prod_{m,y}e^{-i\mathcal{A}^4_{m,n}(\lambda^n_z-\lambda^m_y)},\\
\label{BAE2}
&&\prod_{m,y}e^{-i\mathcal{B}^4_{n,m}(\mu^n_z-\lambda^m_y)}=
(-)^{\delta_{\mu^n}}\prod_{m,y}e^{-i\mathcal{A}^2_{m,n}(\mu^n_z-
\mu^m_y)},
\end{eqnarray}
where
\begin{eqnarray}
&&\mathcal{B}^t_{m,n}(x)=\theta_{\frac{1}{t}(2m+n-1)} +
\theta_{\frac{1}{t}(2m+n-3)} +\theta_{\frac{1}{t}(|2m-n|+1)},\\
&&\mathcal{{A}}^t_{m,n}(x)=\theta_{ \frac{1}{t}(m+n)} + 2
\theta_{\frac{1} {t}(m+n-2)} + \cdots
+2\theta_{\frac{1}{t}(|m-n|)}+\theta_{\frac{1}{t}(|m-n|)},
\end{eqnarray}
and $\delta$ gives the phase shifts $\pi$ or $0$ according to the
numbers of production terms. Taking the logarithm of (\ref{BAE2}),
we have
\begin{eqnarray}\label{BAE22}
&&2\pi I^\lambda_{nz}=
N\theta_{\frac{n}{4}}(\lambda^n_z)
+\sum_{m,y}\mathcal{B}^4_{m,n}(\lambda^n_z-\mu^m_y)-\sum_{m,y}
\mathcal{A}^4_{m,n}
(\lambda^n_z-\lambda^m_y),\nonumber \\ &&2\pi
I^\mu_{my}
=\sum_{n,z}\mathcal{B}^4_{m,n}(\mu^m_y-\lambda^n_z)
-\sum_{n,z}\mathcal{A}^2_{n,m}(\mu^m_y-\mu^n_z),
\end{eqnarray}
where $I^\lambda_{nz}$ and $I^\mu_{my}$ are
the integers or half-integers which
determine the eigenstates. A set of
$\{I^{\lambda}_{nz}, I^\mu_{my}\}$ satisfying
(\ref{BAE22}) gives a highest weight eigenstate of
Hamiltonian (\ref{HSS}). The momentum
(\ref{momentum-P-1}) can be written in terms of
$I^\lambda_{nz}$ and $I^\mu_{my}$
\begin{eqnarray}
\label{momentum-P-2} P &=& 2\pi\frac{1}{N}
\sum_{n,z} (I_{\lambda^n_z} + I_{\lambda^n_z}) \mod 2\pi.
\end{eqnarray}

In the thermodynamic limit, the summations become integrations. Denoting $\eta_n$ and $\sigma_m$ as the densities of $\lambda$
$n$-strings and $\mu$ $m$-strings in the thermodynamic limit, and
$\eta_n^h$ and $\sigma_m^{h}$ as the corresponding densities
of holes, the Bethe ansatz equations read
\begin{eqnarray}
&& \frac{I^{\lambda}}{N}=
\theta_{n/4} (\lambda)
+\sum_{m}\hat{\mathcal{B}}^4_{m,n}*\sigma^m(\lambda)
-\sum_{m}\hat{\mathcal{A}}^4_{m,n}*\eta^m(\lambda),
\nonumber \\
&&\label{BAE3} \frac{I^\mu}{N} =
\sum_{m}\hat{\mathcal{B}}^4_{n,m}*\eta^m(\mu)
-\sum_{m}\hat{\mathcal{A}}^2_{m,n}*\sigma^m(\mu),
\end{eqnarray}
where the densities of $\lambda^n_z$ and $\mu^n_z$ are
denoted as $\eta^n(\lambda_z)$ and $\eta^n(\mu_z)$,
respectively, $I$s are the quantum
numbers, and the operator $*$ is defined by
\begin{eqnarray}
\int a_t(x-y)f(y)\mathrm{d}y = [t]*f(x).
\end{eqnarray}

Taking the differentials of Eq. (\ref{BAE3}), we obtain the integral
form of the Bethe ansatz equations
\begin{eqnarray}
&&\eta^n_h(k) =
a_{{n}/{4}}(k)+\sum_{m=1}\hat
B^4_{m,n}*\sigma^m(k)-\sum_{m=1}\hat
A^4_{m,n}*\eta^m(k),
\nonumber\\
\label{BAE4}
&&\sigma^n_h(k)=\sum_{m=1}\hat B^4_{n,m}*\eta^m(k) -\sum_{m=1}\hat
A^2_{m,n}*\sigma^m(k).
\end{eqnarray}
where $\hat A_{m,n}^t$ and $\hat B_{m,n}^t$ are the integral
operators, $ \hat A_{m,n}^t=a_{(m+n)/t} +2a_{(m+n-2)/t} +\cdots +
2a_{(|m-n|-2)/t} +a_{(|m-n|)/t}$, $ \hat B_{m,n}^t = a_{(2m+n-1)/t}
+ a_{(2m+n-3)/t} +\cdots + a_{(|2m-n|+1)/t}$. In the derivation, we
have used the relation $(\partial/\partial x) \theta_t(x) = 2\pi
a_t(x)$.

At temperature $T$, the Gibbs free energy of the system (1) with an
external magnetic field $h$ reads
\begin{eqnarray}
F=E -h\left(\frac{3}{2}N-M_1-M_2\right)-TS,
\end{eqnarray}
where
\begin{eqnarray}
&&M_1= N\sum_{n} n \int \eta_n(\lambda){\rm d}\lambda, \quad M_2=
N\sum_{m} m \int \sigma_m(\mu) {\rm d} \mu
\\
\label{energy-H-int} &&E = -\frac{9}{2} \pi JN \sum_n \int {\rm d}
\lambda a_{n/4} (\lambda) \eta^n(\lambda) -\frac{99}{8}N ,\\
\label{entro-1} && S= N\sum_{n} \int {\rm d}\lambda
[(\eta_n+\eta_n^h)\ln(\eta_n+\eta_n^h)  -\eta_n\ln\eta_n
-\eta_n^h\ln\eta_n^h] \nonumber\\ && \qquad +N\sum_{m} \int {\rm
d}\mu[(\sigma_m+\sigma_m^h)\ln(\sigma_m+\sigma_m^h)
-\sigma_m\ln\sigma_m-\sigma_m^h\ln\sigma_m^h].
\end{eqnarray}

Minimizing the Gibbs free energy at the thermal equilibrium, we
obtain the following thermodynamic Bethe ansatz equations
\begin{eqnarray}\label{thermal-1}
&& \ln {\tilde \eta}_1 = -\frac{9}{2} \pi J  \frac{G_4(\lambda)} {T}
+ G_4* \ln [(1+\tilde \eta_2)(1+ \tilde \rho^{-1})^{-1}],
\nonumber\\
&& \ln
\tilde \sigma_1 = G_2*\ln  \frac{1+ \tilde \sigma_2} {(1+\tilde
\eta_{1}^{-1}) (1+\tilde \eta_{3}^{-1})} - \frac{G_2}
{G_4}* \ln(1+\tilde \eta_2^{-1}),
\nonumber\\
&& \ln \tilde \eta_{n \in even} = G_4*\ln  \frac{(1+ \tilde
\eta_{n-1}) (1+\tilde \eta_{n+1})}{1+\tilde \sigma_{n/2}^{-1}},
\\
&& \ln \tilde \eta_{n \in odd} = G_4*\ln[ (1+\tilde \eta_{n-1})
(1+\tilde \eta_{n+1}) ],
\nonumber\\
&& \ln \tilde \sigma_m= G_2*\ln  \frac{ (1+\tilde \sigma_{m-1})
(1+\tilde \sigma_{m+1})} {(1+\tilde \eta_{2m-1}^{-1})
(1+\tilde \eta_{2m+1}^{-1})} -\frac{G_2}{G_4}*\ln(1+\tilde \eta_{2m}^{-1}),
\nonumber \\
&&\lim_{n \rightarrow \infty}\frac{\ln \tilde
\eta_n}{n}=\frac{h}{T}, \quad \lim_{m \rightarrow
\infty}\frac{\ln\tilde \sigma_m}{m}=\frac{h}{T}, \nonumber
\end{eqnarray}
where $\tilde \eta_n= \eta_n^h/\eta_n$, $\tilde \sigma_m =
\sigma_m^h/\sigma_m$, $G_n= a_{1/n}/(a_0 +a_{2/n})$ and $a_0\equiv
\delta(x)$. In the derivation, we have used the following relations
\begin{eqnarray}
&&[n+m]=[n][m], \nonumber\\
&&A^{n,m}_t-G_t(A^{n-1,m}_t+A^{n+1,m}_t) =\delta_{m,n},
(n>1), \nonumber\\
&&A^{1,m}_t-G_tA^{2,m}_t =\delta_{1,n}, \nonumber\\
&&B^{m,n}_t-G_t(B^{m,n-1}_t+B^{m,n+1}_t) =\delta_{2m,n},
(n>1), \nonumber\\
&&B^{m,1}_t-G_tB^{m,2}_t=0, \nonumber\\
&&B^{n,m}_t-G_{t/2}(B^{n-1,m}_t+B^{n+1,m}_t) = (\delta_{2m-1,n}
   +\delta_{2m+1,n})G_{t/2}\nonumber\\
&&~~~~~~~~~~~~~~~~~~~~~~~~~~~~~~~~~~~~~~~~~~~~~~~~+\delta_{2m,n}
   G_{t/2}/G_t,(n>1), \nonumber\\
&&B^{1,m}_t-G_{t/2}B^{2,m}_t =(\delta_{1,n}
   +\delta_{2,n})G_{t/2}(n>1)+\delta_{2,n} G_{t/2}/G_t,\\
&&\left[\frac{n+1}{t}\right]A^{n,m}_t -\left[\frac{n}{t}\right]
A^{n+1,m}_t =\left\{\begin{array}{l} 0,(m\leq n)\\-a_{m/t}/G_t,(m\geq n+1)
\end{array}\right., \nonumber\\
&&\left[\frac{n+1}{t}\right]B^{m,n}_t -\left[\frac{n}{t}\right]
B^{m,n+1}_t=\left\{\begin{array}{l} 0,(2m\leq n)\\-a_{2m/t}/G_t
,(2m\geq n+1)   \end{array}\right., \nonumber\\
&&\left[\frac{n+1}{t}\right]B^{n,m}_t -\left[\frac{n}{t}\right]
B^{n+1,m}_t =\left\{\begin{array}{l} 0,(m\leq 2n)\\
   -a_{(m+1)/t},(m=2n+1)\\-a_{m/t}{G_t}
   ,(m\geq 2n+2)    \end{array}\right.. \nonumber
\end{eqnarray}

\section{Ground state}
\label{GS}

In order to obtain the ground state of the system (1), we first
define the dressed energies of $n$-string $\lambda$ and $m$-string
$\mu$
\begin{eqnarray} \label{Dress}
\zeta_n (\lambda) = T \ln \tilde \eta_n(\lambda), \quad
\varsigma_m(\mu) = T\ln \tilde \sigma_n(\mu).
\end{eqnarray}
From Eqs. (\ref{BAE4}) and (\ref{thermal-1}), we find that the
dressed energies should satisfy
\begin{eqnarray}
\label{dress-1} &&\zeta_n(\lambda) = -\frac{9}{2} \pi J
a_{{n}/{4}}(\lambda) + hn -\sum_{m=1}\hat
B^4_{m,n}*\ln[1+e^{-\varsigma_m(\lambda)/T}] \nonumber\\
&&~~~~~~~~~~~~~~~~~~~~~~~~~~~~~~~~~~~~~~~~~~~~~~~~~~~~~~~
+\sum_{m=1}\hat A^4_{m,n}*
\ln[1+e^{-\zeta_m(\lambda)/T}],
\\
&&\varsigma_n(k)=hn- \sum_{m=1}\hat
B^4_{n,m}*\ln[1+e^{-\zeta_m(\lambda) /T}]  +\sum_{m=1}\hat
A^2_{m,n}*\ln[1+e^{-\varsigma_m(\lambda)/T}].
\end{eqnarray}
In the thermodynamic limit, we obtain the following thermodynamic
Bethe ansatz equations for the dressed energy
\begin{eqnarray}\label{dress-2}
&&\zeta_1 = -\frac{9}{2} \pi J G_4(\lambda) + TG_1*\ln
[1+e^{\varsigma_2(\lambda)/T}],  \nonumber\\
&&\varsigma_1 = TG_2* \{\ln[1+e^{\varsigma_2(\lambda)/T}]-
\ln[1+e^{-\zeta_1(\lambda)/T}]-
\ln[1+e^{-\zeta_3(\lambda)/T}]\}\nonumber\\
&& ~~~~~~~~~~~~~~~~~~~~~~~~~~~~~~~~~~~~~~~~~~~~~~~~~~~~~~~
- T\frac{G_2}{G_4}* \ln(1+e^{-\zeta_3(\lambda)/T}),
\nonumber\\
&& \zeta_{n \in {\rm even}} = TG_4*\{\ln[1+e^{\zeta_{n-1}/T} +
\ln[1+e^{\zeta_{n+1}/T}]-\ln[1+e^{-\zeta_{n/2}/T}]\},\\
&& \zeta_{n \in {\rm odd}} = T G_4*\{\ln[1+e^{\zeta_{n-1}/T} +
\ln[1+e^{\zeta_{n+1}/T}]\}, \nonumber\\
&& \varsigma_m= TG_2* [\ln(1+e^{\varsigma_{m-1}/T}) +
\ln(1+e^{\varsigma_{m+1}/T})- \ln(1+e^{-\zeta_{2n+1}/T})
\nonumber\\
&& ~~~~~~~~~~~~~~~~~~~~~~~~~~~~~~
 -\ln(1+e^{-\zeta_{2n-1}/T})]
-T\frac{G_2}{G_4}*\ln(1+e^{-\zeta_{2n}/T}), \nonumber \\
&&\lim_{n \rightarrow \infty}\frac{\zeta_n}{n}=h, \quad \lim_{m
\rightarrow \infty}\frac{\varsigma_m}{m}=h. \nonumber
\end{eqnarray}
The dressed energies can be divided into two parts $\epsilon^+(k)$
and $\epsilon^-(k)$,
\begin{eqnarray}\label{dress-3}
\epsilon^+(k)=\left\{\begin{array}{ll}
\epsilon(k) & \mbox{ if } \epsilon(k)>0,\\ 0 & \mbox{ if } \epsilon(k)<0;
\end{array}\right.~~~~~~
\epsilon^-(k)=\left\{\begin{array}{ll}
\epsilon(k) & \mbox{ if } \epsilon(k)<0,\\ 0 & \mbox{ if } \epsilon(k)>0;
\end{array}\right..
\end{eqnarray}
The ground state string distribution of the system can be obtained
by taking the limit of $T\to 0$ and $h\to 0$. When $T\to 0$ and
$h\to 0$, we have $\ln[1+e^{\epsilon(\lambda)/T}]  =
\ln[1+e^{\epsilon^+(\lambda)/T} e^{\epsilon^-(\lambda)/T}]$, and
$\ln[1+e^{-\epsilon(\lambda)/T}] = \ln[1+e^{-\epsilon^+(\lambda)/T}
e^{-\epsilon^-(\lambda)/T}]$. When $T\to0$, both
$e^{-\epsilon^+(\lambda)/T}$ and $e^{\epsilon^-(\lambda)/T}$ tend to
one. Thus $\ln[1+e^{\epsilon(\lambda)/T}] =
\ln[1+e^{\epsilon^+(\lambda)/T}]$ and
$\ln[1+e^{-\epsilon(\lambda)/T}] =
\ln[1+e^{-\epsilon^-(\lambda)/T}]$.

If $J<0$, the ground state of the system is the ferromagnetic state
$\left|0\right\rangle = \otimes_{j=1}^N\left|3/2\right\rangle$. It
is easy to understand from the eigen energy (\ref{energy-H-2}) for a
$n$-string $\lambda$ give a positive contribution $e_n(\lambda_z)$
to the eigen energy. In the ground state, the total spin along
$z$-direction and the magnetization are $S^z = 3N/2$ and
$\mathfrak{m} = 3/2$, respectively. It is a highest weight
representation of the Yang-algebra (\ref{YBRRTT}). The ground state
energy and momentum are $E = -99JN/8$ and $P = 0$, respectively.
\begin{figure}[h]
\begin{center}
\includegraphics[height=6cm,width=8cm]{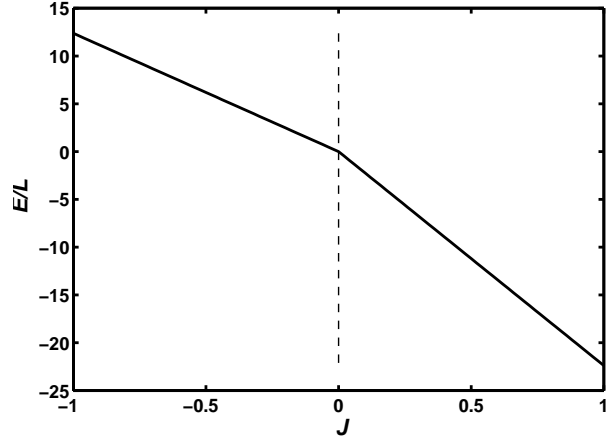}
\end{center}
\caption{The ground state energy of the system. There is a quantum
phase transition at the critical point $J=0$. The system is in the
ferromagnetic phase if $J<0$ and is in the antiferromagnetic phase
if $J>0$.} \label{fig1}
\end{figure}

If $J>0$, we find that some $\lambda$s are real while the others
form 2-strings, and the $\mu$s are real. Such a ground state
configuration is quite different from that of the $SU(4)$ Sutherland
model where there is no string or spin bound state in the ground
state. In the present $SO(5)$ case, part of the spectral parameters
form 2-strings which heavily affect the spin excitations as we shall
show below. To make the ground state energy lowest, all these
strings are filled up and no holes left. This can be understood from
the entropy $S$ of the system. The completely filled string
configurations contribute zero entropy, and make the system in a
most stable sate.

From Eqs. (\ref{BAE4}), the densities $\eta^n_0$ and $\sigma^m_0$
satisfy the following integral equations
\begin{eqnarray}\label{AFg-de-1}
&&\vec \rho_0(\lambda) = \vec g(\lambda) + \vec K* \vec
\rho_0(\lambda),
\end{eqnarray}
where $\vec \rho_0(\lambda) = [\eta_{01} (\lambda),
\eta_{02}(\lambda), \sigma_{01}(\lambda)] ^t$, $\vec g(\lambda) =
[a_{1/4}(\lambda), a_{2/4}(\lambda), 0 ]^t$ and
\begin{eqnarray}
&&\vec K = \left(\begin{array}{rrrrr}
- a_{\frac{1}{2}}&~~& - a_{\frac{3}{4}} -a_{\frac
{1}{4}} &~~& + a_{\frac{1}{2}}
\\
- a_{\frac{3}{4}}-a_{\frac{1}{4}}&&  -a_{1} -2a_{ \frac{1}{2}} &&
a_{\frac{3}{4}} +a_{\frac{1}{4}}
\\
a_{\frac{1}{2}}  &&
a_{\frac{3}{4}}  + a_{\frac{1}{4}} && - a_1
\end{array} \right).
\end{eqnarray}
Taking the inverse of Eq. (\ref{AFg-de-1}), we obtain
\begin{eqnarray}\label{AFg-de-2}
\vec \rho_0 (\lambda)= \vec F* \vec g(\lambda),
\end{eqnarray}
where $\vec F = 1/(1- \vec K)$. The solution of
(\ref{AFg-de-2}) is
\begin{eqnarray}\label{af-des}
&&\eta_{01}(\lambda) = \cosh (\pi \lambda),\nonumber\\
&&\eta_{02} (\lambda) = \frac{1}{18\pi}
\csch(2\pi\lambda) \Big[
4\pi\sqrt{3}\sinh\frac{4\pi\lambda}{3}-12\pi\lambda\Big],\\
&&\sigma_{01} (\lambda) = \cosh(\pi
\lambda/3)/3.\nonumber
\end{eqnarray}
\begin{figure}[h]
\begin{center}
\includegraphics[height=5cm,width=6.5cm]{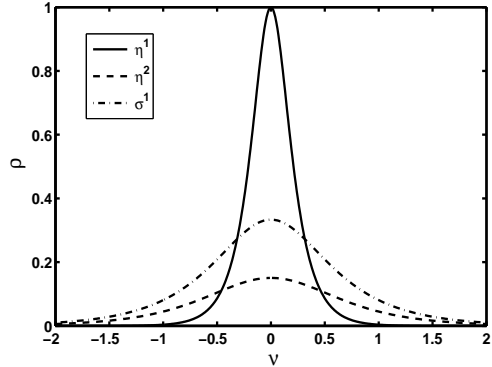}
\end{center}
\caption{The densities of $\eta^1(\lambda)$,
$\eta^2(\lambda)$ and $\sigma^1(\lambda)$ of the
ground state.}
\label{fig2}
\end{figure}
\begin{figure}[h]
\begin{center}
\includegraphics[height=5cm,width=6.5cm]{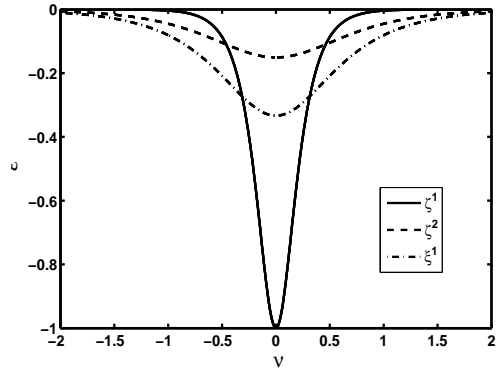}
\end{center}
\caption{The dressed energies $\zeta^1(\lambda)$,
$\zeta^2(\lambda)$ and $\xi^1(\lambda)$ of the ground
state with $J=2/9\pi$.} \label{fig3}
\end{figure}
Because all the density functions are even, the ground state string
configuration
$\{I^1_0(\lambda^{(1)}_i)\},~\{I^2_0(\lambda^{(2)}_j)\},~\{I^1_0(\mu^{(1)}_k)\}$
is symmetric around the origin. Taking the integration, we obtain
the densities of strings $n_{01}^{\eta} = M_1^1/N$, $n_{02}^{\eta} =
M_1^2/N$ and $n_{01}^{\sigma} = M_2^1/N$ as
\begin{eqnarray}
n_{01}^{\eta} =\frac{1}{2}, ~~~ n_{02}^{\eta} =\frac{1}{4},
~~~ n_{01}^{\sigma} =\frac{1}{2},
\end{eqnarray}
After some derivations, we find $M_1=N$ and $M_2=N/2$, which mean
that the total spin of the ground state is zero, $S=3N/2 -M_1-M_2=
0$. Thus the ground state is a spin singlet state.  The ground state
eigen energy and the momentum are
\begin{eqnarray}
\label{AF-g-E} &&E_0^A = -\frac{9}{2}\pi JNF(0) -
\frac{99}{8}JN,\hspace{10pt}  F(\lambda)= a_{1/4}*\eta_{01}(\lambda)
+a_{1/2}*\eta_{02}(\lambda),
\\
\label{AF-g-P}
&&P_0^A =   2\pi\frac{1}{N} \sum_{n,z} (I_{\lambda^n_z} +
I_{\lambda^n_z}) \mod 2\pi=0.
\end{eqnarray}
From Eq. (\ref{dress-1}), the ground state dressed energies satisfy
the following equations
\begin{eqnarray}\label{AF-dr}
&&\vec \epsilon (\lambda) = -\frac{9}{2}\pi J \vec F*
\vec g(\lambda) = -\frac{9}{2} \pi J\vec \rho_0(\lambda),
\end{eqnarray}
where $\vec \epsilon (\lambda) = [\zeta_1(\lambda),
\zeta_2(\lambda), \varsigma_1(\lambda) ]^T$. Recalling the
definition of the dressed energy $\epsilon  = T\ln(\rho^h/\rho)$, we
see that the dressed energies are negative in the limit of $T\to 0$,
which means that the corresponding strings are completely filled.

\section{Elementary excitations}
\label{Exc}

Based on the ground state configuration, the elementary excitations
of the system can be studied exactly. For the ferromagnetic case
($J<0$), elementary excitations are spin waves with dispersion
relation $\Delta E = -18(J/n)\cos^2 (n\Delta P/2)$. The excitations
in anti-ferromagnetic sector ($J>0$) are somehow complicated. In the
language of Bethe ansatz, these can be described by the changes of
the string distributions. These excitations are very different from
those of the $SU(4)$ model, for the ground state configuration
contains 2-strings. In fact, the spin excitations can be described
by adding some holes or high strings into the ground state
configuration.

The  holes and extra strings lead to redistributions of $\lambda$s
and $\mu$s. Formally, the extra strings contribute nothing to the
energy because the contribution of such strings is exactly canceled
by the rearrangement of the ground state distribution though they do
contribute to the spin quanta carried by the excitations.

The excited states can be determined by the integral
Bethe ansatz equations with holes and high strings in
the $\lambda^{1}$, $\lambda^{2}$ and
$\mu^{1}$ sectors
\begin{eqnarray}\label{AFed}
&&\vec \rho(\lambda)  = \vec F* [\vec g(\lambda) - \vec
\rho^h(\lambda)],
\end{eqnarray}
where $\vec \rho (\lambda) = [\eta_1(\lambda), \eta_2(\lambda),
\sigma_1(\lambda)]^T$ and $\vec \rho^h(\lambda)= [\eta^h_1(\lambda),
\eta^h_2(\lambda), \sigma^h_1(\lambda)]^T$. In the thermodynamic limit the density of holes are
\begin{eqnarray}
&&\eta^h_a(\nu) = \sum_{i=1} ^{m_a} \frac{1}{N} \delta(\nu-\nu^a_i),
\end{eqnarray}
where $m_{1,2,3}$ represent the numbers of holes in real $\lambda$
sea, in 2-string $\lambda$ sea and in real $\mu$ sea, $\nu^1_i$,
$\nu^2_j$ and $\nu^3_k$ are the positions of the corresponding
charges and holes. The excitations lead to the redistributions of
densities $\vec \rho_0(\lambda)$,
\begin{eqnarray}
&& \vec \rho(\lambda) + \vec \rho^h(\lambda) -\vec \rho_0(k) = \vec
K* [\vec \rho(\lambda) - \vec \rho_0(k)].
\end{eqnarray}
Denoting the charges of the densities as $  \Delta \vec \rho(k)
=\vec \rho(\lambda) - \vec \rho_0(\lambda)$, which satisfies
\begin{eqnarray}
&& \Delta \vec \rho(\lambda)  = -\vec F* \vec \rho_h(\lambda).
\end{eqnarray}
Form Eq. (\ref{energy-H-int}), the excited energy is
\begin{eqnarray}\label{energy-int-eci}
\Delta E &=&E- E^A_0 =-N \sum_{a} \left.\epsilon_{a} * \rho^h_a
(\lambda) \right|_{\lambda=0} = -\sum_{a=1}^3 \sum_{i=1}^{m_a}
\epsilon_a(\nu_a^i).
\end{eqnarray}
Thus the excited energies are the summation of dressed energies
carried by the holes with a inverse sign. The excited momentum is
\begin{eqnarray}\label{momentum-int-eci}
&&\Delta P = P-P^A_0= \sum_{a=1}^3\sum_{i=a}^{m_a} \int_0^{\nu_a^i}
\rho_{0a}(\lambda) {\rm d} \lambda \mod 2\pi.
\end{eqnarray}
The spin quanta carried by the spin excitation is
\begin{eqnarray}\label{spin-int-eci}
S=\frac{3}{2}m_2+2m_3+\sum_{l\geq3}(2-l)m_{\lambda^{(l)}} +
\sum_{t\geq2}(1-t)m_{\mu^{(t)}},
\end{eqnarray}
where $m_{\lambda^{(l)}}$ and $m_{\mu^{(t)}}$ are the numbers of
$\lambda$ $l$-strings and the $\mu$ $t$-strings formed in the
excitations, respectively.

\begin{figure}[t]
\begin{center}
\includegraphics[height=5cm,width=6.5cm]{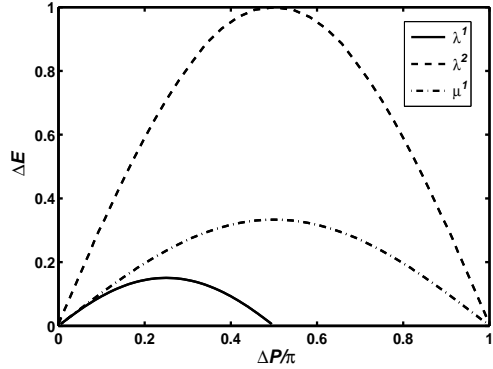}
\end{center}
\caption{The single-hole excitations of the system.
Here $J=2/9\pi$, $\Delta E$ and $P$ are the energy and
the momentum carried by a single hole, respectively.
The dotted dashed line is the single-hole excitation
of real $\lambda$. The solid line is that of $\lambda$
2-string and the dashed line is that of real $\mu$.}
\label{fig4}
\end{figure}

Because the energies, momenta and spins of the holes are additive,
the thermodynamic behaviors of the system are mainly determined by
the dispersion relations of the individual holes, which are shown in
Fig.\ref{fig3}. From the Fig.\ref{fig3}, we find that the single
$\lambda$ 2-string hole carries the lowest energy with spin $3/2$,
which is named as heavy spinon here. These heavy spinons dominate
the low temperature thermodynamics of the system. Surprisingly, the
holes in the real $\lambda$ sector carry zero spin, corresponding to
a new kind of neutral spin excitations. The spin quanta carried by
each $\mu$ hole are 2. Different from the $SU(4)$ model, the
$\lambda$ 2-string heavy spinons cover one quarter of the Brilliouin
zone (see Fig. \ref{fig4}). This might be detected by the neutron
scattering methods.

The numbers of holes and strings added are not independent but
satisfy some constraints determined by the Bethe ansatz equations
(\ref{BAE4}),
\begin{eqnarray}
&&\Delta M_1^1=-m_1+\frac{1}{2}m_2, \nonumber\\&&\Delta M_1^2 =
\frac{1}{2}m_1-\frac{3}{4}m_2-\frac{1}{2}m_3 -\sum_{l\geq3}
m_{\lambda^{(l)}}, \\ &&\Delta M_2^1=-\frac{1}{2}m_2 -m_3
-\sum_{t\geq2} m_{\mu^{(t)}},\nonumber
\end{eqnarray}
where $m_{1,2,3}$ are the numbers of holes in the real $\lambda$,
2-string $\lambda$ and real $\mu$-sea, $m_{\lambda^{(l)}}$ and
$m_{\mu^{(t)}}$ are the number of $l$-string in the rapidity
$\lambda$ and $t$-string in the rapidity $\mu$, respectively. Thus
$\Delta M_{1,2}^l$ are some integers which indicate the number
changes of $\lambda, \mu$ $l$-strings. For convenience, we define
$\Delta M_a=\sum_i \Delta M_a^i$ and denote these excitations as $
{\mathbf m} = [(m_1,m_2,m_3), (m_\lambda^l,m_\lambda^{l'},\cdots)$,
$(m_\mu^t, m_\mu^{t'}, \cdots)]$. As we mentioned above, the excited
momenta, excited energies and the spins are additive, $\Delta
P_{\mathbf m+ \mathbf m'} = \Delta P_{\mathbf m}+ \Delta P_{\mathbf
m'}$, $\Delta E_{\mathbf m+ \mathbf m'} =\Delta E_{\mathbf m}+\Delta
E_{\mathbf m'}$ and $S^z_{\mathbf m+ \mathbf m'} =S^z_{\mathbf
m}+S^z_{\mathbf m'}$. The number changes of $\Delta M_a^b$ are also
additive
\begin{eqnarray}
{\Delta M_a^b}_{\mathbf m+ \mathbf m'} =  {\Delta M_a^b}_{\mathbf m}
+ {\Delta M_a^b}_{\mathbf m'},~~
{\Delta M_a}_{\mathbf m+ \mathbf m'} =  {\Delta M_a}_{\mathbf m}
+ {\Delta M_a}_{\mathbf m'}.
\end{eqnarray}

Some possible hole configurations are listed in Table
\ref{table}. We denote $[(a,b,c), (0, 0, \cdots)$,
$(0,0,\cdots)] = [(a,b,c)]$ for short. We
find that the excitations $\mathbf m^0_1 = [(2,0,0)]$,
$\mathbf m^0_2=[(1,0,1)]$, $\mathbf m^0_3=[(1,2,0)]$,
$\mathbf m^0_4=[(0,0,2)]$, $\mathbf m^0_5=[(0,2,1)]$
and $\mathbf m^0_6= [(0,4,0)]$ are the basic
excitations. The additional single high strings ${\bf
m^\lambda}_l$ are not independent. They can form the
excitations with ${\bf m}^0_i$, such as ${\bf m}_1$,
${\bf m}_2$, ${\bf m}_3$, ${\bf m}_4$ and ${\bf m}_5$.

\begin{table}[t]
\begin{center}
\caption{\label{table} Some possible configurations of the
excitations. Here, $\Delta M_a=\sum_i \Delta M_a^i$ and ${\bf
m}^0_i$ are basic excitations. ${\bf m^\lambda}_l$ and ${\bf
m}^\mu_t$ are the high string excitations. Some possible elementary
excitations ${\bf m}_1 = {\bf m}^0_2 + {\bf m}^\mu_2$, ${\bf m}_2 =
{\bf m}^0_3 + {\bf m}^\lambda_2$, ${\bf m}_3 = {\bf m}^0_3 + 2{\bf
m}^\lambda_3$,
${\bf m}_4 = {\bf m}^0_5 + {\bf m}^\lambda_6+2{\bf m}^\mu_2$ and
${\bf m}_5 = {\bf m}^0_5 + {\bf m}^\lambda_6+{\bf m}^\mu_3$ are
shown. Here $a\times b$ means the number of $b$-strings added is $a$.}
\begin{tabular}{l|ccccc|ccc|cc|c}
\hline\hline $\bf m$&$m_1$&$m_2$&$m_3$
&$m_{\lambda^{(l)}}$&$m_{\mu^{(t)}}$& $\Delta M_1^1$&$\Delta M_1^2$&
$\Delta M_2^1$&$\Delta M_1$& $\Delta M_2$&$\Delta S^z$\\ \hline
${\bf m}^0_1$&$2$&$0$&$0$&   $0$&$0$   &$-2$&$ 1$&$ 0$   & $0$& $0$ &$0$\\
${\bf m}^0_2$&$1$&$0$&$1$&   $0$&$0$   &$-1$&$ 0$&$-1$   &$-1$&$-1$ &$2$\\
${\bf m}^0_3$&$1$&$2$&$0$&   $0$&$0$   &$ 0$&$-1$&$-1$   &$-2$&$-1$ &$3$\\
${\bf m}^0_4$&$0$&$2$&$1$&   $0$&$0$   &$ 1$&$-2$&$-2$   &$-3$&$-2$ &$5$\\
${\bf m}^0_5$&$0$&$4$&$0$&   $0$&$0$   &$ 2$&$-3$&$-2$   &$-4$&$-2$ &$6$\\
${\bf m}^0_6$&$0$&$0$&$2$&   $0$&$0$   &$ 0$&$-1$&$-2$   &$-2$&$-2$ &$4$\\
\hline
${\bf m}^\lambda_{l}$&$0$&$0$&$0$&$1\times l$&$0$&$0$&$-1$&$0$&$l-1$&$0$&$-l$\\
${\bf m}^\mu_{t}$&$0$&$0$&$0$&$0$&$1\times t$&$0$&$0$&$-1$&$t-1$&$0$&$-t$\\
\hline
$\bf m_1$&$1$&$0$&$1$&$0$&$1\times2$       &$-1$&$ 0$&$-2$ &$-1$&$0 $&$1$\\
$\bf m_2$&$1$&$2$&$0$&$1\times4$&       $0$&$ 0$&$-2$&$-1$ &$ 0$&$-1$&$1$\\
$\bf m_3$&$1$&$2$&$0$&$2\times3$&       $0$&$ 0$&$-3$&$-1$ &$ 0$&$-1$&$1$\\
$\bf m_4$&$0$&$4$&$0$&$1\times6$&$2\times2$&$ 2$&$-4$&$-4$ &$0$&$ 0$ &$0$\\
$\bf m_5$&$0$&$4$&$0$&$1\times6$&$1\times3$&$ 2$&$-4$&$-3$ &$0$&$ 0$ &$0$\\
\hline\hline
\end{tabular}
\end{center}
\end{table}

\begin{figure}[t]
\begin{center}
\includegraphics[height=7.5cm,width=5.2in]{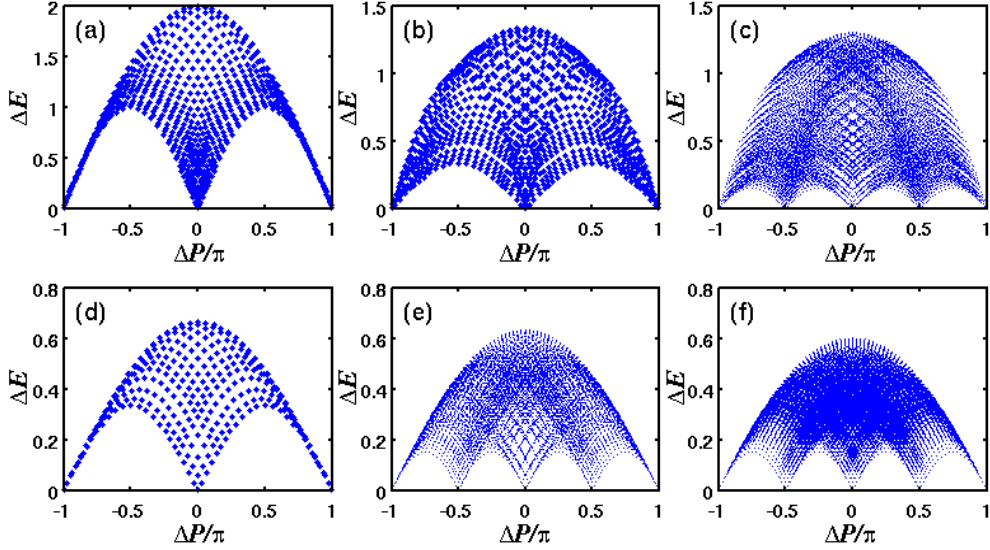}
\end{center}
\caption{The low-lying excitations of the system. Here $J=2\pi/9$,
$\Delta E$ and $\Delta P$ are the energy and the momentum carried by
the excitation, respectively. (a) ${\bf m}_1^0=[(200)]$; (b) ${\bf
m}_2^0=[(101)]$; (c) ${\bf m}_3^0=[(120)]$; (d) ${\bf
m}_6^0=[(002)]$; (e) ${\bf m}_4^0=[(021)]$; (f) ${\bf
m}_5^0=[(040)]$.} \label{fig5}
\end{figure}

The low-lying excitations ${\bf m}^0_i$ are shown in Fig.\ref{fig2}.
The simplest spin excitation is a real $\lambda$ hole-pair ${\bf
m}^0_1$ that is two-neutral spinon excitation, corresponding to the
two domain walls of a single excited domain (Fig. \ref{fig5}(a)).
The $\lambda$ 2-string hole pair can not exist independently. They
must be associated with a neutral spinon, i. e. ${\bf m}^0_3$ (Fig.
\ref{fig5}(c)). Accompanied by a real $\mu$ hole, the $\lambda$
2-string hole pair is also a possible excitation ${\bf m}^0_5$ (Fig.
\ref{fig5} (e)). Further, if we add a $\lambda$ 4-string into a
2-string hole pair and a real $\lambda$ hole in the $\lambda$-sea,
the total spin of this excitation is 1 (${\bf m}_2$ in Table.
\ref{table}). In this case each of the $\lambda$ 2-string holes
carries a spin $1/2$. Such a dressed hole is quite similar to the
ordinary spinon, and it is named as dressed spinon here. Four
$\lambda$ 2-string holes ${\bf m}^0_5$ may exist independently (Fig.
\ref{fig5}). If we put further one $\lambda$ 6-string and one $\mu$
3-strings into this four hole configuration (${\bf m}_5$ in Table.
\ref{table}), we obtain the $SO(5)$ spin singlet excitation. The
simplest excitation in the $\mu$ sector is a pair of real $\mu$
holes ${\bf m}^0_6$ (Fig. \ref{fig5}(d)). This excitation is quite
similar to a real $\lambda$ hole pair but each of the real $\mu$
hole carries a spin $2$. Joint pair of a real $\lambda$ hole and a
real $\mu$ hole ${\bf m}^0_2$ may also happen as shown in Table.
\ref{table}. Other kinds of spin excitations such as ${\bf m}_1$,
${\bf m}_3$ and ${\bf m}_4$ can be constructed similarly.

\section{Conclusion}
\label{C}

In conclusion, we propose an integrable spin-3/2 chain model with
$SO(5)$ symmetry. By using the nested quantum inverse scattering
method, we obtain the exact solutions of the system. Different from
the $SU(4)$ integrable spin chain, there only exist three conserved
quantities. Based on the exact solutions, the ground state and
thermodynamic properties of the system are discussed. Several new
kinds of spin excitations such as the neutral spin excitations,
heavy spinons and dressed spinons are found.

\section*{Acknowledgement}

This work was supported by the NSFC, the Knowledge Innovation
Project of CAS, and the National Program for Basic Research of MOST.

\end{document}